\documentclass[10pt,final,journal,tbtags,twoside,romanappendices]{IEEEtran}

\usepackage{epsfig, graphics, color}
\usepackage{rotating, times, xspace, cite, amssymb}
\usepackage[cmex10]{amsmath}
\DeclareMathAlphabet{\mathpzc}{OT1}{pzc}{m}{it}
\DeclareMathAlphabet{\mathss}{OT1}{cmss}{m}{sl}
\newlength{\figsize}
\renewcommand{\IEEEQED}{\IEEEQEDopen}
\newtheorem{definition}{Definition}
\newtheorem{theorem}{Theorem}

\newtheorem{remark}{Remark}
\allowdisplaybreaks[2]
\newcommand{\s}[1]{{\mathcal{#1}}}	
\renewcommand{\v}[1]{{\mathbf{#1}}}	
\newcommand{\vg}[1]{\boldsymbol{#1}}	

\providecommand{\Min}[1]{\min\braces{#1}}

\newcommand{\mmax}[1]{\bracket{#1}^+}
\newcommand{\braces}[1]{{\left\{ {#1}\right\}}}
\newcommand{\paren}[1]{\left({#1}\right)}
\newcommand{\bracket}[1]{{\left [{#1}\right ]}}

\newcommand{\prob}[2][]{\mathrm{P}_{#1} \{#2\}}
\newcommand{\Normal}[1]{\mathcal{N}\paren{#1}}
\newcommand{\isnormal}[1]{\sim \Normal{#1}}
\newcommand{\del}{\partial}
\newcommand{\Lag}{\mathcal{L}}

\newcommand{\ital}[1]{{\em #1}}
\renewcommand{\bold}[1]{{\bf #1}}
\newcommand{\tozero}{\rightarrow 0}
\newcommand{\toinf}{\rightarrow \infty}
\newcommand{\toone}{\rightarrow 1}
\newcommand{\onehalf}{\frac{1}{2}}
\newcommand{\nvar}{\sigma^2}
\newcommand{\e}{\ensuremath{\epsilon}\xspace}
\newcommand{\n}{\ensuremath{n}\xspace }
\newcommand{\Perr}[1][]{\ensuremath{P_e^{#1}}}
\newcommand{\sumton}[1][i]{\sum_{#1=1}^n}
\newcommand{\ssum}{{\textstyle \sum}}
\newcommand{\Markov}[3]{#1 \rightarrow #2 \rightarrow #3}
\newcommand{\Popt}{P^*}
\newcommand{\Pmax}{\bar{P}}
\newcommand{\Pmopt}{\v{P}^*}
\newcommand{\Pmmax}{\bar{\v{P}}}

\newcommand{\ninv}[1][1]{\frac{#1}{n}}
\newcommand{\muv}{\vg{\mu}}

\newcommand{\Nm}{\v{N}}
\newcommand{\Pm}{\v{P}}
\newcommand{\Rm}{\v{R}}
\newcommand{\Wm}{\v{W}}
\newcommand{\Xm}{\v{X}}
\newcommand{\Zm}{\v{Z}}
\newcommand{\Ym}{\v{Y}}

\newcommand{\Ks}{\s{K}}

\newcommand{\Ps}{\s{P}}
\newcommand{\Ws}{\s{W}}
\newcommand{\Ss}{\s{S}}
\newcommand{\Ssc}{{\Ss^c}}
\newcommand{\efrac}[3][\;]{\frac{#1 #2 #1}{#1 #3 #1}} 
\newcommand{\mnormal}[1]{{\displaystyle #1}}

\newcommand{\Cs}{\ensuremath{\s{C}}}
\newcommand{\Gs}{\ensuremath{\s{G}}}
\newcommand{\channel}[1]{{\scriptscriptstyle \text{#1}}}
\newcommand{\Mch}{\channel{M}}
\newcommand{\Wch}{\channel{W}}

\newcommand{\CM}{C^{\Mch}}
\newcommand{\CW}{C^{\Wch}}
\newcommand{\CWs}{\tilde{C}^{\Wch}}
\newcommand{\secsign}[1]{{#1}^s}
\newcommand{\pubsign}[1]{{#1}^o}
\newcommand{\xsign}[1]{{#1}^x}
\newcommand{\tsign}[1]{{#1}^t}

\newcommand{\Wmh}{\hat \Wm}
\newcommand{\Wsec}{\secsign{W}}
\newcommand{\Wpub}{\pubsign{W}}
\newcommand{\Wx}{\xsign{W}}
\newcommand{\Wmsec}{\secsign{\Wm}}
\newcommand{\Wmpub}{\pubsign{\Wm}}

\newcommand{\Wmsech}{\secsign{\Wmh}}
\newcommand{\Wmpubh}{\pubsign{\Wmh}}

\newcommand{\Wssec}{\secsign{\Ws}}
\newcommand{\Wspub}{\pubsign{\Ws}}

\newcommand{\Rsec}{\secsign{R}}
\newcommand{\Rpub}{\pubsign{R}}
\newcommand{\Rx}{\xsign{R}}
\newcommand{\Rt}{\tsign{R}}
\newcommand{\Msec}{\secsign{M}}
\newcommand{\Mpub}{\pubsign{M}}
\newcommand{\Mx}{\xsign{M}}
\newcommand{\Mt}{\tsign{M}}
\newcommand{\Xc}{\mathfrak{X}}

\newcommand{\subsc}[2]{#1_{\scriptscriptstyle {#2}}}
\renewcommand{\Cs}[1][]{\ensuremath{\s{C}^{\scriptscriptstyle \text{#1}}}\xspace}
\renewcommand{\Gs}[1][]{\ensuremath{\s{G}^{\scriptscriptstyle \text{#1}}}\xspace}

\newcommand{\NM}{\Nm_\Mch}
\newcommand{\NW}{\Nm_\Wch}

\newcommand{\hM}{h^{\Mch}}
\newcommand{\hW}{h^{\Wch}}

\newcommand{\secrecy}{perfect secrecy\xspace}

\newcommand{\CsMAC}{\Cs[\text{MA}]}
\newcommand{\GsMAC}{\Gs[\text{MA}]}
\newcommand{\GsMACS}{\Gs[\text{MA-SUP}]}
\newcommand{\GsMACT}{\Gs[\text{MA-TDMA}]}
\newcommand{\CsTW}{\Cs[\text{TW}]}
\newcommand{\GsTW}{\Gs[\text{TW}]}
\newcommand{\Wmsecp}{\secsign{\acute \Wm}}
\newcommand{\Rmsec}{\secsign{\Rm}}

\newcommand{\Rmsecp}{\secsign{\acute \Rm}}
\newcommand{\Wssecp}{\secsign{\acute \Ws}}
\newcommand{\Wspubp}{\pubsign{\acute \Ws}}
\newcommand{\NMt}{\subsc{\tilde{\Nm}}{\Mch}}
\newcommand{\NWt}{\subsc{\tilde{\Nm}}{\Wch}}
\newcommand{\XmS}{\Xm_\Sigma}
\newcommand{\Xcs}{\secsign{\Xc}}
\newcommand{\Xcp}{\pubsign{\Xc}}
\newcommand{\Xcx}{\xsign{\Xc}}
\newcommand{\lambdas}{\secsign{\lambda}}
\newcommand{\lambdap}{\pubsign{\lambda}}
\newcommand{\lambdax}{\xsign{\lambda}}
\newcommand{\h}{\mathss{h}}

\newcommand{\Rsum}[1][]{\ensuremath{R_{sum}^{\scriptscriptstyle \text{#1}}}\xspace}
\newcommand{\RsumMACS}{\Rsum[MA-SUP]}

\newcommand{\RsumTW}{\Rsum[TW]}
\newcommand{\RsumMACJ}{\Rsum[SUP-MA-CJ]}
\newcommand{\RsumTWJ}{\Rsum[TW-CJ]}
\newcommand{\hMA}{\phi}
\newcommand{\hdMA}{\dot{\hMA}}
\newcommand{\hdjMA}{\hdMA^{\scriptscriptstyle (j)}}
\newcommand{\hdjoptMA}{\hdMA_\Ks^{\scriptscriptstyle (j)}(\Pmopt)}

\newcommand{\hoptMA}{\hMA_\Ks(\Pmopt)}

\newcommand{\hJMA}{\Phi_\Ts(\Pm)}
\newcommand{\hJoptMA}{\Phi_\Ts(\Pmopt)}
\newcommand{\hJoptoptMA}{\Phi_\Tsopt(\Pmopt)}

\newcommand{\hTW}{\psi}
\newcommand{\hdTW}{\dot{\hTW}}
\newcommand{\hdjTW}{\hdTW^{\scriptscriptstyle (j)}}
\newcommand{\hdjoptTW}{\hdTW_\Ks^{\scriptscriptstyle (j)}(\Pmopt)}

\newcommand{\hoptTW}{\hTW_\Ks(\Pmopt)}

\newcommand{\Ts}{\s{T}}
\newcommand{\jc}{{j^c}}

\newcommand{\Tsc}{{\Ts^c}}
\newcommand{\Psmax}{\bar \Ps}

\newcommand{\Tsalt}{{\s{U}}}
\newcommand{\Tsaltc}{{\Tsalt^c}}
\newcommand{\Tsopt}{{\Ts^*}}
\newcommand{\Tsoptc}{{\Ts^{*c}}}
\newcommand{\Palt}{Q}
\newcommand{\Pmalt}{\v{Q}}

\title{The General Gaussian Multiple Access and Two-Way Wire-Tap Channels:\\
Achievable Rates and Cooperative Jamming}
\author{
Ender~Tekin,~\IEEEmembership{Student~Member,~IEEE} and
Aylin~Yener,~\IEEEmembership{Member,~IEEE}
\thanks{Mauscript received February 16, 2007; revised September 30, 2007. This work has been supported by NSF grant CCF-0514813 ``Multiuser Wireless Security" and DARPA ITMANET Program grant W911NF-07-1-0028.  This work was presented in part in the 2006 Allerton Conference on Communications, Control, and Computing, \cite{tekin:ALLERTON06}, and 2007 International Symposium on Information Theory, \cite{tekin:ISIT07}.}%
\thanks{The authors are with the Department of Electrical Engineering at the Pennsylvania State University, University Park, PA 16802 (email: tekin@psu.edu, yener@ee.psu.edu).}
\thanks{Digital Object Identifier}}

\begin{document}
\thispagestyle{headings}
\maketitle
\markboth{IEEE Transactions on Information Theory}
	{Tekin and Yener: The General Gaussian Multiple Access and Two-Way Wire-Tap Channels}

\begin{abstract}
The General Gaussian Multiple Access Wire-Tap Channel (GGMAC-WT) and the Gaussian Two-Way Wire-Tap Channel (GTW-WT) are considered.  In the GGMAC-WT, multiple users communicate with an intended receiver in the presence of an eavesdropper who receives their signals through another GMAC. In the GTW-WT, two users communicate with each other over a common Gaussian channel, with an eavesdropper listening through a GMAC.  A secrecy measure that is suitable for this multi-terminal environment is defined, and achievable secrecy rate regions are found for both channels. For both cases, the power allocations maximizing the achievable secrecy sum-rate are determined. It is seen that the optimum policy may prevent some terminals from transmission in order to preserve the secrecy of the system.  Inspired by this construct, a new scheme, \ital{cooperative jamming}, is proposed, where users who are prevented from transmitting according to the secrecy sum-rate maximizing power allocation policy ``jam" the eavesdropper, thereby helping the remaining users.  This scheme is shown to increase the achievable secrecy sum-rate.  Overall, our results show that in multiple-access scenarios, users can help each other to collectively achieve positive secrecy rates.  In other words, cooperation among users can be invaluable for achieving secrecy for the system.
\end{abstract}

\section{Introduction}
\label{sec:intro}
\IEEEPARstart{G}{aussian} multiple-access channels and two-way channels are two of the earliest channels that were considered in the literature.  The multiple-access channel capacity region was determined in \cite{ahlswede:multiway,liao:multiaccess}.  The two-way channel was initially examined by Shannon, \cite{shannon:twoway}, where he found inner and outer bounds for the general two-way channel, and determined the capacity region for some special cases.  In \cite{dueck:twoway}, it was shown that the inner bound found by Shannon was not tight in general.  The capacity region of the Gaussian two-way channel was found by Han in \cite{han:twoway}.  A related, somewhat more general case called two-user channels was studied in \cite{ahlswede:twouser, sato:twouser}.  For a comprehensive review of these channels, the reader is referred to \cite{elgamal-cover:MUIT}.

A rigorous analysis of information theoretic secrecy was first given by Shannon in \cite{shannon:secrecy}. In this work, Shannon showed that to achieve \ital{perfect secrecy} in communications, which is equivalent to providing no information to an enemy cryptanalyst, the conditional probability of the \ital{cryptogram given a message} must be independent of the actual transmitted message.  In other words, the \ital{a posteriori} probability of a message must be equivalent to its \ital{a priori} probability.

In \cite{wyner:wiretap}, Wyner applied this concept to the discrete memoryless channel.  He defined the wire-tap channel, where there is a wire-tapper who has access to a degraded version of the intended receiver's signal.  Using the normalized conditional entropy $\Delta$ of the transmitted message given the received signal at the wire-tapper as the secrecy measure, he found the region of all possible $(R,\Delta)$ pairs, and the existence of a \ital{secrecy capacity}, $C_s$, the rate up to which it is possible to limit the rate of information transmitted to the wire-tapper to arbitrarily small values.

In \cite{hellman-carleial:wiretap}, it was shown that for Wyner's wire-tap channel, it is possible to send several low-rate messages, each completely protected from the wire-tapper individually, and use the channel at close to capacity.  However, if any of the messages are available to the wire-tapper, the secrecy of the rest may also be compromised.  Reference \cite{leung-hellman:gaussianwiretap} extended Wyner's results in \cite{wyner:wiretap} and Carleial and Hellman's results in \cite{hellman-carleial:wiretap} to Gaussian channels.  The seminal work by Csisz\'ar and K\"orner, \cite{csiszar-korner:confbroadcast}, generalized Wyner's results to ``less noisy" and ``more capable" channels. Furthermore, it examined sending common information to both the receiver and the wire-tapper, while maintaining the secrecy of some private information that is communicated to the intended receiver only.  Reference \cite{maurer-wolf:weaktostrongsecrecy} suggested that the secrecy constraint developed by Wyner needed to be strengthened, since it constrains the rate of information leaked to the wire-tapper, rather than the total information, and the information of interest might be in this small amount.  It was then shown that the results of \cite{wyner:wiretap, csiszar-korner:confbroadcast} can be extended to ``strong" secrecy constraints for discrete channels, where the limit is on the total leaked information rather than just the rate, with no loss in achievable rates, \cite{maurer-wolf:weaktostrongsecrecy}.

In the past two decades, common randomness has emerged as a valuable resource for secret key generation,  \cite{maurer:secretkeypublicdiscussion, bennettetal:genprivacy}.  In \cite{maurer:secretkeypublicdiscussion}, it was shown that the existence of a ``public" feedback channel can enable the two parties to be able to generate a secret key even when the wire-tap capacity is zero.  References \cite{ahlswede-csiszar:CR1} and \cite{ahlswede-csiszar:CR2} examined the secret key capacity and \ital{common randomness} capacity, for several channels.   These results also benefit from \cite{maurer-wolf:weaktostrongsecrecy} to provide ``strong" secret key capacities. Maurer also examined the case of active adversaries, where the wire-tapper has read/write access to the channel in \cite{maurer-wolf:SecretKey1}--\nocite{maurer-wolf:SecretKey2} \cite{maurer-wolf:SecretKey3}.  The secret key generation problem was investigated from a multi-party point of view in \cite{venkatesan-anantharam:CRcap-pair} and \cite{venkatesan-anantharam:CRcap-network}. Notably, Csisz\'ar and Narayan considered the case of multiple terminals where a number of terminals try to distill a secret key and a subset of these terminals can act as helper terminals to the rest in \cite{csiszar-narayan:CRhelper}, \cite{csiszar-narayan:secrecycap-multi}.

Recently, several new models have emerged, examining secrecy for parallel channels \cite{yamamoto:secretsharing, yamamoto:secretsharinggaussian}, relay channels \cite{oohama:relaywiretap}, and fading channels \cite{barros:fadingwiretap, blochetal:wirelesssec1}.  Fading and parallel channels were examined together in \cite{liang:Allerton06a, zang:parallelsecrecy}. Broadcast and interference channels with confidential messages were considered in \cite{liuetal:IBCconf}.  References \cite{liang:genMACconfPAP, liuetal:MACconf} examined the multiple access channel with confidential messages where two transmitters try to keep their messages secret from each other while communicating with a common receiver. In \cite{liang:genMACconfPAP}, an achievable region was found in general, and the capacity region was found for some special cases.  MIMO channels were considered in \cite{khisti:MISOME, shafieeetal:MIMOWT221}.

In \cite{tekin:ASILOMAR05, tekin:ISIT06, tekin:IT06a, tekin:ALLERTON06}, we investigated multiple access channels where transmitters communicate with an intended receiver in the presence of an external wire-tapper from whom the messages must be kept confidential.  In \cite{tekin:ASILOMAR05, tekin:ISIT06, tekin:IT06a}, we considered the case where the wire-tapper gets a degraded version of a GMAC signal, and defined two separate secrecy measures extending Wyner's measure to multi-user channels to reflect the level of trust the network may have in each node.  Achievable rate regions were found for different secrecy constraints, and it was shown that the secrecy sum-capacity can be achieved using Gaussian inputs and stochastic encoders.  In addition, TDMA was shown to also achieve the secrecy sum-capacity.

In this paper, we consider the General Gaussian Multiple Access Wire-Tap Channel (GGMAC-WT) and the Gaussian Two-Way Wire-Tap Channel (GTW-WT), both of which are of interest in wireless communications as they correspond to the case where a single physical channel is utilized by multiple transmitters, such as in an ad-hoc network.  We consider an external \ital{eavesdropper}\footnote{Even though we faithfully follow Wyner's terminology in naming the channels, admittedly in wireless system models, \ital{eavesdropper} is a more appropriate term for the adversary.} that receives the transmitters' signals through a general Gaussian multiple access channel (GGMAC) in both system models.  We utilize a suitable secrecy constraint which is the normalized conditional entropy of the transmitted secret messages given the eavesdropper's signal, corresponding to the ``collective secrecy" constraints used in \cite{tekin:IT06a}. We show that satisfying this constraint implies the secrecy of the messages for all users.  In both scenarios, transmitters are assumed to have one secret and one open message to transmit.  This is different from \cite{tekin:IT06a} in that the secrecy rates are not constrained to be at least a fixed portion of the overall rates.  We find an achievable \ital{secrecy rate region}, where users can communicate with arbitrarily small probability of error with the intended receiver under \ital{\secrecy} from the eavesdropper, which corresponds to the result of \cite{tekin:IT06a} for the degraded case.  We note that, in accordance with the recent literature, when we use the term \secrecy, we are referring to ``weak" secrecy, where the \ital{rate} of information leaked to the adversary is limited. As such, this can be thought of as ``almost perfect secrecy".  We also find the sum-rate maximizing power allocations for the general case, which is more interesting from a practical point of view.  It is seen that as long as the users are not \ital{single-user decodable} at the eavesdropper, a secrecy-rate trade off is possible between the users.  Next, we show that a non-transmitting user can help increase the secrecy capacity for a transmitting user by effectively ``jamming" the eavesdropper, and even enable secret communications that would not be possible in a single-user scenario.  We term this new scheme \ital{cooperative jamming}.  The GTW-WT is shown to be especially useful for secret communications, as the multiple-access nature of the channel hurts the eavesdropper without affecting the communication rate.  This is due to the fact that the transmitted messages of each user essentially help hide the other user's secret messages, and reduce the extra randomness needed in wire-tap channels to confuse the eavesdropper.

The rest of the paper is organized as follows: Section \ref{sec:system} describes the system model for the GGMAC-WT and GTW-WT and the problem statement.  Section \ref{sec:ach} describes the general achievable rates for the GGMAC-WT and GTW-WT.  Sections \ref{sec:summax} and \ref{sec:jam} give the sum-secrecy rate maximizing power allocations, and the achievable rates with cooperative jamming.  Section \ref{sec:results} gives our numerical results followed by our conclusions and future work in Section \ref{sec:conclusion}.

\section{System Model and Problem Statement}
\label{sec:system}
We consider $K$ users communicating in the presence of an eavesdropper who has the same capabilities.  Each transmitter $k \in \Ks \triangleq \{1,\dotsc,K\}$ has two messages, $\Wsec_k$ which is secret and $\Wpub_k$ which is open\footnote{We would like to stress that \ital{open} is not the same as \ital{public}, i.e., we do not impose a decodability constraint for the open messages at the eavesdropper.}, from two sets of equally likely messages $\Wssec_k=\{1, \dotsc, \Msec_k\}$, $\Wspub_k=\{1, \dotsc, \Mpub_k\}$.  Let $\Wm_k=(\Wsec_k,\Wpub_k)$, $\Ws_k=\Wssec_k \times \Wspub_k$, $M_k=\Msec_k \Mpub_k$, $\Wmpub_\Ss=\{\Wpub_k\}_{k \in\Ss}$, and $\Wmsec_\Ss = \{\Wsec_k\}_{k \in \Ss}$.  The messages are encoded using $(2^{nR_k},n)$ codes into $\{\tilde X_k^n(\Wm_k)\}$, where $R_k=\ninv \log_2 M_k=\ninv \log_2 \Msec_k + \ninv \log_2 \Mpub_k = \Rsec_k + \Rpub_k$. 
The encoded messages $\{\tilde \Xm_k\}=\{\tilde X_k^n\}$ are then transmitted.  We assume the channel parameters are universally known, and that the eavesdropper also has knowledge of the codebooks and the coding scheme.  In other words, there is no shared secret.  The two channels we consider in this paper are described next.

\subsection{The General Gaussian Multiple-Access Wire-Tap Channel}
\label{sec:sysMAC}
\begin{figure}[t]
\begin{center}
\includegraphics[width=\figsize, angle=0]{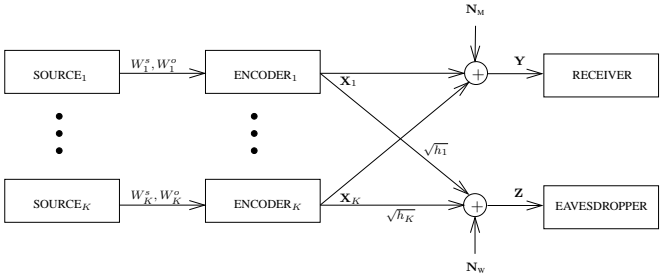}
\caption{\small The standardized GMAC-WT system model.}
\label{fig:gmacwt2}
\end{center}
\vspace{-.2in}
\end{figure}

This is a scenario where the users communicate with a common base station in the presence of an eavesdropper, where both channels are modeled as Gaussian multiple-access channels as shown in Figure \ref{fig:gmacwt2}.  The intended receiver and the wire-tapper receive $\tilde \Ym=\tilde Y^n$ and $\tilde \Zm=\tilde Z^n$, respectively.  The receiver decodes $\tilde \Ym$ to get an estimate of the transmitted messages, $\Wmsech_\Ks,\Wmpubh_\Ks$.  We would like to communicate with the receiver with arbitrarily low probability of error, while keeping the wire-tapper (eavesdropper) ignorant of the secret messages, $\Wmsec_\Ks$.  The signals at the intended receiver and the wire-tapper are given by
\begin{subequations}
\label{eqn:MAC}
\begin{align}
\tilde \Ym &= \sum_{k=1}^K \sqrt{\hM_k} \tilde \Xm_k + \NMt \\
\tilde \Zm &= \sum_{k=1}^K \sqrt{\hW_k} \tilde \Xm_k + \NWt
\end{align}
\end{subequations}
where $\NMt,\NWt$ are the AWGN, $\tilde \Xm_k$ is the transmitted codeword of user $k$, and $\hM_k,\hW_k$ are the channel gains of user $k$ to the intended receiver (\ital{main} channel, M), and the eavesdropper (\ital{wire-tap} channel, W), respectively.  Each component of $\NMt \isnormal{0,\nvar_\Mch}$ and $\NWt \isnormal{0,\nvar_\Wch}$.
We also assume the following transmit power constraints:
\begin{equation}
\label{eqn:powerconstraints}
\ninv \sumton{\tilde X_{ki}^2} \le \tilde{\Pmax}_k, \; k=1,\dotsc,K.
\end{equation}

Similar to the scaling transformation to obtain the standard form of the interference channel, \cite{carleial:interference}, we can represent any GMAC-WT by an equivalent standard form, \cite{tekin:IT06a}:
\begin{subequations}
\label{eqn:MACstd}
\begin{align}
\Ym &= \sum_{k=1}^K \Xm_k + \NM \\
\Zm &= \sum_{k=1}^K  \sqrt{h_k} \Xm_k + \NW
\end{align}
\end{subequations}
where, for each $k$, 
\begin{itemize}
\item the codewords are scaled to get $\Xm_k = \sqrt{\frac{\hM_k}{\nvar_\Mch}}\tilde \Xm_k$;
\item The new power constraints are $\Pmax_k = \frac{\hM_k}{\nvar_\Mch} \tilde \Pmax_k$;
\item The wiretapper's new channel gains are $h_k = \frac{\hW_k \nvar_\Mch}{\hM_k \nvar_\Wch}$;
\item The noises are normalized to get $\NM = \frac{\NMt}{\nvar_\Mch}$ and $\NW = \frac{\NWt}{\nvar_\Wch}$.
\end{itemize}

We can show that the eavesdropper gets a stochastically degraded version of the receiver's signal if $h_1=\dotsc=h_K \equiv h < 1$.  We considered this special case in \cite{tekin:ISIT06,tekin:IT06a}.

\subsection{The Gaussian Two-Way Wire-Tap Channel} 
\begin{figure}[t]
\begin{center}
\includegraphics[width=\figsize, angle=0]{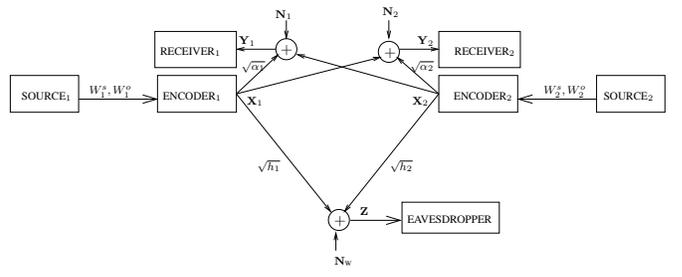}
\caption{\small The standardized GTW-WT system model.}
\label{fig:tw}
\end{center}
\vspace{-.2in}
\end{figure}

In this scenario, two transmitter/receiver pairs communicate with each other over a common channel.  Each receiver $k=1,2$ gets $\tilde \Ym_k=\tilde Y_k^n$ and the eavesdropper gets $\tilde \Zm=\tilde Z^n$.  Receiver $k$ decodes $\tilde \Ym_k$ to get an estimate of the transmitted messages of the other user.  The users would like to communicate the open and secret messages with arbitrarily low probability of error, while maintaining secrecy of the secret messages.
The signals at the intended receiver and the wiretapper are given by
\begin{subequations}
\label{eqn:TW}
\begin{align}
\tilde \Ym_1 &= \tilde \Xm_1 + \sqrt{\hM_2} \tilde \Xm_2 + \tilde \Nm_1 \\
\tilde \Ym_2 &= \sqrt{\hM_1} \tilde \Xm_1 + \tilde \Xm_2 + \tilde \Nm_2 \\
\tilde \Zm &= \sqrt{\hW_1} \tilde \Xm_1 + \sqrt{\hW_2} \tilde \Xm_2 + \NWt
\end{align}
\end{subequations}
where $\tilde \Nm_k \isnormal{0,\nvar_k}$ and $\NWt \isnormal{0,\nvar_\Wch}$.
We also assume the same power constraints given in \eqref{eqn:powerconstraints} (with $K=2$), and again use an equivalent standard form as illustrated in Figure \ref{fig:tw}:
\begin{subequations}
\label{eqn:TWstd}
\begin{align}
\Ym_1 &= \sqrt{\alpha_1} \Xm_1 +\Xm_2 + \Nm_1 \\
\Ym_2 &= \Xm_1 +\sqrt{\alpha_2} \Xm_2 + \Nm_2 \\
\Zm &= \sqrt{h_1} \Xm_1 + \sqrt{h_2} \Xm_2 + \NW
\end{align}
\end{subequations}
where 
\begin{itemize}
\item the codewords $\{\tilde \Xm\}$ are scaled to get $\Xm_1 = \sqrt{\frac{\hM_1}{\nvar_2}} \tilde \Xm_1$ and $\Xm_2 = \sqrt{\frac{\hM_2}{\nvar_1}} \tilde \Xm_2$;
\item the maximum powers are scaled to get $\Pmax_1 = \frac{\hM_1}{\nvar_2} \tilde \Pmax_1$ and $\Pmax_2 = \frac{\hM_2}{\nvar_1} \tilde \Pmax_2$;
\item the transmitters' new channel gains are given by $\alpha_1 = \frac{\nvar_2}{\hM_1 \nvar_1}$ and $\alpha_2 = \frac{\nvar_1}{\hM_2 \nvar_2}$;
\item the wiretapper's new channel gains are given by $h_1 = \frac{\hW_1 \nvar_2}{\hM_1 \nvar_\Wch}$ and $h_2 = \frac{\hW_2 \nvar_1}{\hM_2 \nvar_\Wch}$;
\item the noises are normalized by $\Nm_k = \frac{1}{\nvar_k} \tilde \Nm_k, \, k=1,2$ and $\NW = \frac{1}{\nvar_\Wch}\NWt$.
\end{itemize}

\subsection{Preliminary Definitions}
\label{sec:predef}
In this section, we present some useful preliminary definitions including the secrecy constraint we will use.  In particular, the secrecy constraint we used is the ``collective secrecy constraint" we defined in \cite{tekin:ASILOMAR05,tekin:IT06a}, and is suitable for the multi-access nature of the systems of interest.
\begin{definition}[Collective secrecy constraint]
We use the \ital{normalized joint conditional entropy} of the transmitted messages given the eavesdropper's received signal as our secrecy constraint, i.e.,
\begin{equation}
\Delta_\Ss \triangleq \frac{H(\Wmsec_\Ss|\Zm)}{H(\Wmsec_\Ss)}
\end{equation}
for any set $\Ss \subseteq \Ks$ of users.  For perfect secrecy of all transmitted secret messages, we would like 
\begin{equation}
\label{eqn:DeltaKdef}
\Delta_\Ks = \frac{H(\Wmsec_\Ks|\Zm)}{H(\Wmsec_\Ks)} \toone.
\end{equation}
\end{definition}

Assume $\Delta_\Ks \ge 1-\e$ for some arbitrarily small $\e$ as required.  Then,
\begin{align}
H(\Wmsec_\Ks|\Zm) &\ge H(\Wmsec_\Ks) -\e H(\Wmsec_\Ks) \\
H(\Wmsec_\Ss|\Zm) &\ge H(\Wmsec_\Ss) + H(\Wmsec_\Ssc|\Wmsec_\Ss) \notag \\
	&\qquad -\e H(\Wmsec_\Ks) - H(\Wmsec_\Ssc|\Wmsec_\Ss,\Zm) \\
	&\ge H(\Wmsec_\Ss) - \e H(\Wmsec_\Ks) \\
\Delta_\Ss &\ge 1-\e'
\end{align}
where $\e' \triangleq \frac{H(\Wmsec_\Ks)}{H(\Wmsec_\Ss)}\e \tozero$ as $\e \tozero$.  If $H(\Wmsec_\Ss)=0$, then we define $\Delta_\Ss=1$.  Thus, the perfect secrecy of the system implies the perfect secrecy of any group of users, guaranteeing that when the system is secure, so is each individual user.

\begin{definition}[Achievable rates]
\label{def:achrate}
Let $\Rm_k =(\Rsec_k,\Rpub_k)$.  The rate vector $\Rm=\paren{\Rm_1,\dotsc,\Rm_K}$ is said to be \ital{achievable} if for any given $\e>0$ there exists a code of sufficient length \n such that
\begin{subequations}
\label{eqn:achdef}
\begin{align}
\ninv \log \Msec_k &\ge \Rsec_k - \e, \qquad k=1,\dotsc,K\\
\ninv \log \Mpub_k &\ge \Rpub_k - \e, \qquad k=1,\dotsc,K
\end{align}
and
\begin{equation}
\Perr = \frac{1}{\prod_{k=1}^K M_k} \sum_{\Wm \in {\displaystyle \times}_{k=1}^K \Ws_k}	
	\hspace{-.2in} \prob{\Wmh \neq \Wm |\Wm \text{ sent}} \le \e
\end{equation}
is the average probability of error.  In addition, we need
\begin{equation}
\Delta_\Ks \ge 1-\e
\end{equation}
\end{subequations}
where $\Delta_\Ks$ denotes our secrecy constraint and is defined in \eqref{eqn:DeltaKdef}.  We will call the set of all achievable rates, the \ital{secrecy-capacity region}, and denote it $\CsMAC$ for the GGMAC-WT, and $\CsTW$ for the GTW-WT, respectively.
\end{definition}

Before we state our results, we also define the following notation which will be used extensively in the rest of this paper:
\vspace{-\abovedisplayshortskip}
\begin{align}
\label{eqn:mmax}
\mmax{\xi} &\triangleq \max \bracket{\xi,0} \\
\label{eqn:capM}
\CM_\Ss(\Pm) &\triangleq \onehalf \log \paren{1+\ssum_{k \in \Ss} P_k}, \quad \Ss \subseteq \Ks\\
\label{eqn:capW}
\CW_\Ss(\Pm) &\triangleq \onehalf \log \paren{1+\ssum_{k \in \Ss} h_k P_k}, \quad \Ss \subseteq \Ks\\
\label{eqn:capWs}
\CWs_\Ss(\Pm) &\triangleq \onehalf \log \paren{1+\frac{\ssum_{k \in \Ss} h_k P_k}
	{1+\ssum_{k \in \Ssc} h_k P_k}}, \quad \Ss \subseteq \Ks\\
\label{eqn:Pset}
\Ps &\triangleq \braces{\Pm: 	0 \le P_k \le \Pmax_k, \, \forall k}\\
\label{eqn:Pmmax}
\Pmmax &\triangleq \braces{\Pmax_1,\dotsc,\Pmax_K}
\end{align}

Lastly, we informally call the $k$th user \ital{strong} if $h_k \le 1$, and \ital{weak} if $h_k >1$.  This is a way of indicating whether the intended receiver or the wiretapper is at a more of an advantage concerning that user, and is equivalent to stating whether the single-user secrecy capacity of that user is positive or zero.  We later extend this concept to refer to users who can achieve positive secrecy rates and those who cannot.  In addition, we will say that a user is \ital{single-user decodable} if its rate is such that it can be decoded by treating the other user as noise.  A user group $\Ss$ is single-user decodable by the eavesdropper if $\CM_\Ss(\Pm) \le \CWs_\Ss(\Pm)$.  Our achievable rates cannot guarantee secrecy for such a group of users.
\section{Achievable Secrecy Rate Regions}
\label{sec:ach}

\subsection{The General Gaussian Multiple Access Wire-Tap Channel}
\label{sec:MAC}

In this section, we present our main results for the GGMAC-WT.  We first define two separate regions and then give an achievable region:

\begin{definition}[GGMAC-WT Superposition Region] Let $X_k \isnormal{0,P_k}$ for all $k$.  Then, the superposition region, $\GsMACS$, is given by
\begin{equation}
\hspace{-\multlinegap}%
\begin{split}
\GsMACS(\Pm) \triangleq 
	\Bigl\{ \Rm \colon \hspace{-1in} &\\
	&\sum_{k \in \Ss} \paren{\Rsec_k+\Rpub_k} \le I(\Xm_\Ss;Y|\Xm_\Ssc),
		\quad \forall \Ss \subseteq \Ks \\
	&\sum_{k \in \Ss} \Rsec_k \le \mmax{I(\Xm_\Ss;Y|\Xm_\Ssc)-I(\Xm_\Ss;Z)}, 
		\; \forall \Ss \subseteq \Ks
	\Bigr\}
\end{split}
\end{equation}
which can be written as
\begin{equation}
\label{eqn:MACachP}
\hspace{-\multlinegap}%
\begin{split}
\GsMACS(\Pm)= 
	\biggl \{ \Rm \colon \hspace{-1.0in} & \\
	&\sum_{k \in \Ss} \paren{\Rsec_k{+}\Rpub_k} \le \onehalf \log 
		\paren{1{+}\ssum_{k \in \Ss} P_k},	\quad \forall \Ss {\subseteq} \Ks \\
	&\sum_{k \in \Ss} \Rsec_k \le \onehalf 
		\biggl [ \log \biggl( 1{+}\sum_{k \in \Ss} P_k \biggr) \\
		&\hspace{.7in} {-}\log\paren{1{+}\frac{\sum_{k \in \Ss} h_k P_k}
			{1{+}\sum_{k \in \Ssc} h_k P_k}} 
		\biggr ]^+ \!\!,\; \forall \Ss {\subseteq} \Ks
	\biggr \}.
\end{split}
\end{equation}
\end{definition}

\begin{definition}[GGMAC-WT TDMA Region] Let $\{\alpha_k\}$ be such that $0 \le \alpha_k \le 1$ for all $k$ and $\sum_{k=1}^K \alpha_k=1$. Let $X_k \isnormal{0,P_k/\alpha_k}$ for all $k$.  Then, the TDMA region, $\GsMACT$, is given by
\begin{equation}
\hspace{-\multlinegap}%
\begin{split}
\GsMACT(\Pm,\vg{\alpha}) \triangleq 
	\Bigl \{ \Rm \colon \hspace{-1.2in} & \\
	&\Rsec_k + \Rpub_k \le \alpha_k I(X_k;Y|\Xm_{k^c}),
		\quad \forall k {\in} \Ks \\
	&\Rsec_k \le \alpha_k \mmax{I(X_k;Y|\Xm_{k^c})-I(X_k;Z|\Xm_{k^c})} \!,
		\; \forall k {\in} \Ks
	\Bigr \}
\end{split}
\end{equation}
which is equivalent to
\begin{equation}
\hspace{-\multlinegap}%
\begin{split}
\label{eqn:MACachT}
\GsMACT(\Pm,\vg{\alpha}) =  
	\biggl\{ \Rm \colon \hspace{-1.2in} &\\
	&\Rsec_k + \Rpub_k \le \frac{\alpha_k}{2}\log \paren{1+\frac{P_k}{\alpha_k}},
		\quad \forall k {\in} \Ks \\
	&\Rsec_k \le \! \frac{\alpha_k}{2} \biggl[ \log \!\paren{1{+}\frac{P_k}{\alpha_k}} 
		{-} \log \!\paren{1{+}\frac{h_k P_k}{\alpha_k}}
		\biggr]^+ \!\!, \forall k {\in} \Ks
	\biggr \}.\!
\end{split}
\end{equation}
\end{definition}

\begin{remark}
The superposition and TDMA regions can also be written as follows:
\begin{gather}
\begin{split}
\GsMACS(\Pm) = \Bigl \{ \Rm \colon \hspace{-1in} &\\
	&\sum_{k \in \Ss} \paren{\Rsec_k+\Rpub_k} \le \CM_\Ss(\Pm),
		\quad \forall \Ss \subseteq \Ks \\
	&\sum_{k \in \Ss} \Rsec_k \le \mmax{\CM_\Ss(\Pm)-\CWs_\Ss(\Pm) }, 
		\quad \forall \Ss \subseteq \Ks
	\Bigr \}
\end{split}
\\
\begin{split}
\GsMACT(\Pm,\vg{\alpha}) = \Bigl \{ \Rm \colon \hspace{-1.2in} &\\
	&\Rsec_k + \Rpub_k \le \alpha_k \CM_k \paren{\frac{\Pmax_k}{\alpha_k}},
		\quad \forall k \in \Ks \\
	&\Rsec_k \le \alpha_k \mmax{\CM_k \paren{\frac{\Pmax_k}{\alpha_k}} 
		{-} \CW_k\paren{\frac{\Pmax_k}{\alpha_k}}}\!, \, \forall k \in \Ks
	\Bigr \}
\end{split}
\end{gather}
in accordance with the definitions in \eqref{eqn:capM}--\eqref{eqn:capWs}.
\end{remark}

\begin{theorem}
\label{thm:MACach}
The rate region given below is achievable for the GGMAC-WT:
\begin{multline}
\GsMAC = \text{convex closure of } \\
	\paren{\bigcup_{\Pm \in \Ps} \GsMACS(\Pm)} \bigcup 
	\Biggl( \bigcup_{\substack{\v{0} \le \vg{\alpha} \le \v{1} \\ \Sigma_k \alpha_k=1}}	
		\GsMACT(\Pmmax,\v{\alpha}) \Biggr).
\label{eqn:MACach}
\end{multline}
\end{theorem}

\begin{IEEEproof}
We first show that the superposition encoding rate region given in \eqref{eqn:MACachP} for a fixed power allocation is achievable.  Consider the following coding scheme for rates $\Rm \in \GsMACS(\Pm)$ for some $\Pm \in \Ps$:

\noindent \bold{Superposition Encoding Scheme:} For each user $k$, consider the following scheme:
\begin{IEEEenumerate}
\item	Generate $3$ codebooks $\Xcs_k,\Xcp_k$ and $\Xcx_k$.  $\Xcs_k$ consists of $\Msec_k$	codewords, each component of which is drawn from $\Normal{0,\lambdas_k P_k -\varepsilon}$. Codebook $\Xcp_k$ has $\Mpub_k$ codewords with each component randomly drawn from $\Normal{0,\lambdap_k P_k-\varepsilon}$ and $\Xcx_k$ has $\Mx_k$ codewords with each component randomly drawn from $\Normal{0,\lambdax_k P_k-\varepsilon}$ where $\varepsilon$ is an arbitrarily small number to ensure that the power constraints on the codewords are satisfied with high probability and $\lambdas_k+\lambdap_k+\lambdax_k=1$. Define $\Rx_k=\ninv \log \Mx_k$ and $\Mt_k=\Msec_k \Mpub_k \Mx_k$.

\item To transmit message $\Wm_k =(\Wsec_k, \Wpub_k) \in \Wssec_k \times \Wspub_k$, user $k$ finds the $2$ codewords corresponding to components of $\Wm_k$ and also uniformly chooses a codeword $\Wx_k$ from $\Xcx_k$. User $k$ then adds all these codewords and transmits the resulting codeword, $\Xm_k$, so that it actually transmits one of $\Mt_k$ codewords. Let $\Rt_k = \ninv \log \Mt_k = \Rpub_k+\Rsec_k+\Rx_k$.  Note that since all codewords are chosen uniformly, user $k$ essentially transmits one of $\Mpub_k \Mx_k$ codewords at random for each message $\Wsec_k$, and its overall rate of transmission is $\Rt_k$.
\end{IEEEenumerate}

Specifically, we choose the rates to satisfy
\begin{align}
\hspace{-\multlinegap}%
&\sum_{k \in \Ss} \paren{\Rsec_k{+}\Rpub_k{+}\Rx_k} 
	\le \onehalf \log \biggl(1{+}\sum_{k \in \Ss} P_k \biggr),\, \forall \Ss \subseteq \Ks\\
&\sum_{k \in \Ss} \paren{\Rpub_k{+}\Rx_k} 
	\le \onehalf \log \biggl (1{+}\sum_{k \in \Ss} h_k P_k \biggr), 
	\quad \forall \Ss{\subseteq}\Ks, \notag \\
	&\hspace{1.8in} \text{with equality if } \Ss=\Ks\\
&\sum_{k \in \Ss} \Rsec_k
	\le \onehalf \bigg[\log \biggl( 1{+}\sum_{k \in \Ss} P_k \biggr) \notag \\
	&\hspace{.85in} {-} \log\biggl( 1{+}\frac{\sum_{k \in \Ss} h_k P_k} 
	{1{+}\sum_{k \in \Ssc} h_k P_k} \biggr) \bigg]^+ \!,
	\, \forall \Ss {\subseteq} \Ks \!
\end{align}
which we can also write as:
\begin{align}
&\ssum_{k \in \Ss} \paren{\Rsec_k{+}\Rpub_k{+}\Rx_k} \le \CM_\Ss, 
	\quad \forall \Ss {\subseteq} \Ks
	\label{eqn:achcolM} \\
&\ssum_{k \in \Ss} \paren{\Rpub_k{+}\Rx_k} \le \CW_\Ss, 
	\; \forall \Ss {\subseteq} \Ks, \, \text{with equality if } \Ss{=}\Ks
	\label{eqn:MACRran} \\
&\ssum_{k \in \Ss} \Rsec_k \le \bigl[ \CM_\Ss{-}\CWs_\Ss \bigr]^+\!, 
	\quad \forall \Ss {\subseteq} \Ks.
	\label{eqn:MACcolsec}
\end{align}

Note that if \eqref{eqn:MACcolsec} is zero for a group of users, we cannot achieve secrecy for those users.  When $\Ss=\Ks$, if the sum-capacity of the main channel is less than that of the eavesdropper channel, i.e., $\CM_\Ks \le \CW_\Ks$, secrecy is not possible for the system. Assume this quantity is positive.  To ensure that we can mutually satisfy both \eqref{eqn:MACcolsec}, \eqref{eqn:MACRran}, we can reclassify some open messages as secret.  Clearly, if we can guarantee secrecy for a larger set of messages, secrecy is achieved for the original messages.  From the first set of conditions in \eqref{eqn:MACach} and the GMAC coding theorem, \cite{cover-thomas:IT}, with high probability the receiver can decode the codewords with low probability of error. To show the secrecy condition in \eqref{eqn:achdef}, first note that, the coding scheme described is equivalent to each user $k$ selecting one of $\Msec_k$ messages, and sending a uniformly chosen codeword from among $\Mpub_k \Mx_k$ codewords for each.  Define $\XmS=\sum_{k=1}^K\sqrt{h_k} \Xm_k$, and we have
\begin{align}
H(\Wmsec_\Ks|\Zm) 
	&= H(\Wmsec_\Ks)-I(\Wmsec_\Ks;\Zm) \\
	&= H(\Wmsec_\Ks) - I(\Wmsec_\Ks;\Zm) + I(\Wmsec_\Ks;\Zm|\XmS) \label{eqn:achprf0c} \\
	&= H(\Wmsec_\Ks) - \h(\Zm) + \h(\Zm|\Wmsec_\Ks) \notag \\
		& \hspace{.7in} + \h(\Zm|\XmS) - \h(\Zm|\Wmsec_\Ks,\XmS) \\
	&\label{eqn:achprf1}= H(\Wmsec_\Ks) - I(\XmS;\Zm)+I(\XmS;\Zm|\Wmsec_\Ks)
\end{align}
where we used $\Markov{\Wmsec}{\XmS}{\Zm}$, and thus we have $\h(\Zm|\Wmsec,\XmS)=\h(\Zm|\XmS)$ to get \eqref{eqn:achprf1}.  We will consider the two terms individually.  First, we have the trivial bound due to channel capacity:
\begin{equation}
\label{eqn:achprf2}
I(\XmS;\Zm) \le n \CW_\Ks(\Pm).
\end{equation}

Now write
\begin{equation}
\label{eqn:achprf3}
I(\XmS;\Zm|\Wmsec_\Ks) = H(\XmS|\Wmsec_\Ks)-H(\XmS|\Wmsec_\Ks,\Zm).
\end{equation}
Since user $k$ independently sends one of $\Mpub_k \Mx_k$ codewords equally likely for each secret message,
\begin{align}
H(\XmS|\Wmsec_\Ks) &= \log \paren{\prod_{k=1}^K (\Mpub_k \Mx_k)}\\
	&= n \paren{\sum_{k=1}^K (\Rpub_k + \Rx_k)} \\
	&= n \CW_\Ks(\Pm).
	\label{eqn:achprf4a}
\end{align}

We can also write
\begin{equation}
\label{eqn:achprf4b}
H(\XmS|\Wmsec_\Ks,\Zm) \le n\delta_n
\end{equation}
where $\delta_n \tozero$ as $n \toinf$ since, with high probability, the eavesdropper can decode $\XmS$ given $\Wmsec_\Ks$ due to \eqref{eqn:MACRran} and code generation.
Using \eqref{eqn:achprf2}, \eqref{eqn:achprf3}, \eqref{eqn:achprf4a} and \eqref{eqn:achprf4b} in \eqref{eqn:achprf1}, we get
\begin{align}
H(\Wmsec_\Ks|\Zm) 
	&\ge H(\Wmsec_\Ks)-n \CW_\Ks(\Pm)+n \CW_\Ks(\Pm)-n\delta_n\\ 
	&= H(\Wmsec_\Ks)-n\delta_n.
		\label{eqn:achprfdone}
\end{align}

Now, let us consider the TDMA region given in \eqref{eqn:MACachT}.  This region is obtained when users who can achieve single-user secrecy use a single-user wire-tap code as in \cite{leung-hellman:gaussianwiretap} in a TDMA schedule, where the time-share of each user $k$ is given by $0 \le \alpha_k \le 1$ and $\sum_{k=1}^K \alpha_k=1$.  A transmitter $k$ who can achieve secrecy, i.e., having $h_k<1$, tranmits for $\alpha_k$ portion of the time when all other users are silent, using $\frac{\Pmax_k}{\alpha_k}$ power, satisfying its average power constraint over the TDMA time-frame.  This approach was used in \cite{tekin:IT06a} to achieve secrecy sum-capacity for individual constraints.  When the channel is degraded, i.e., $h_k = h$ for all $k \in \Ks$, then for collective constraints the TDMA region is seen to be a subset of the superposition region.  However, this is not necessarily true for the general case, and by time-sharing between the two schemes we can generally achieve a larger achievable region, given in \eqref{eqn:MACach}.
\end{IEEEproof}

We remark that it is possible to further divide the ``open" messages to get more sets of ``private" messages which are also perfectly secret, i.e., if we let $\Wspub_k = \Wssecp_k \times \Wspubp_k, \, \forall k$, then as long as we impose the same restrictions on $\Rmsecp$ as $\Rmsec$, we can achieve perfect secrecy of $\Wmsecp$, as in \cite{leung-hellman:gaussianwiretap}.  However, this does not mean that we have perfect secrecy at channel capacity, as the secrecy sub-codes carry information about each other.
\begin{figure}[t]
\centering
\includegraphics[width=\figsize,height=.8\figsize, angle=0]{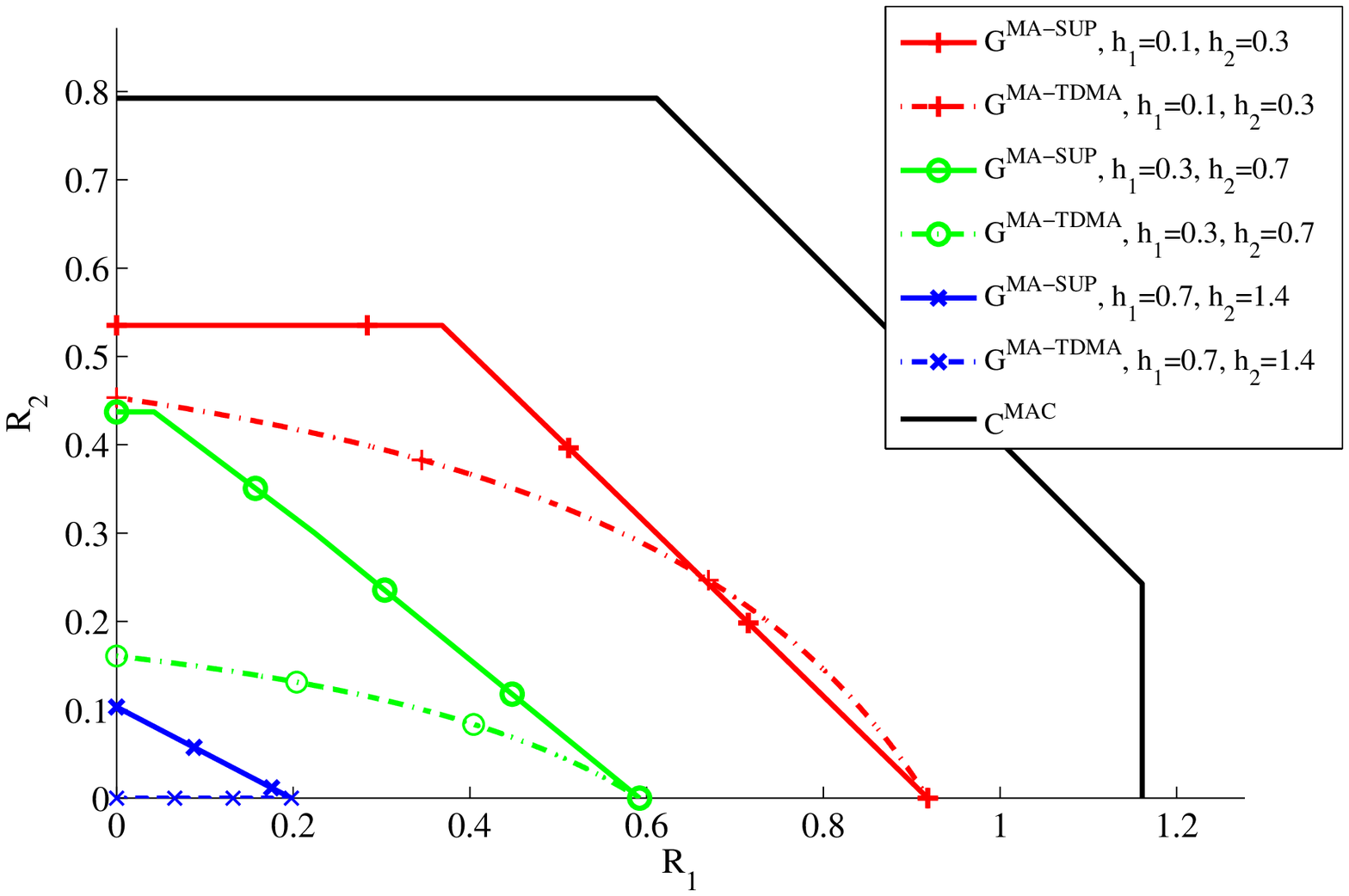}
\caption{\small GGMAC-WT achievable regions for different channel parameters, $\GsMAC(P_1=4, P_2=2)$.}
\label{fig:ggmacwtreg}
\vspace{-.2in}
\end{figure}

\begin{figure}[t]
\centering
\includegraphics[width=\figsize,height=.8\figsize,angle=0]{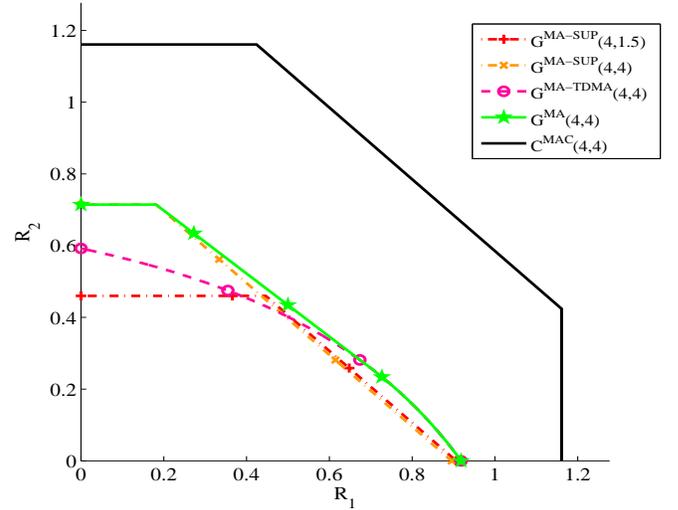}
\caption{\small GGMAC-WT achievable secrecy region when $\Pmax_1=4, \Pmax_2=4, h_1=.1, h_2=.3$.}
\label{fig:ggmacwtUreg}
\vspace{-.18in}
\end{figure}

Observe that even for $K=2$ users, a rate point in this region is four dimensional, and hence cannot be accurately drawn.  We can instead focus on the \ital{secrecy rate region}, the region of all achievable $\Rmsec$. The sub-regions $\GsMACS,\GsMACT$ are shown for different channel gains in Figure \ref{fig:ggmacwtreg} for fixed transmit powers, and $K=2$ users. Figure \ref{fig:ggmacwtUreg} represents how these regions change with different transmit powers when the channel gains are fixed. For the case shown, we need the convex hull operation, as the achievable region is a combination of different superposition and TDMA regions.  Note also that the main extra condition for the superposition region is on the \ital{total extra randomness} added. As a result, it is possible for ``stronger" users to help ``weak" users by contributing more to the necessary extra number of codewords, which is the sum-capacity of the eavesdropper. Such a weak user only has to make sure that it is not single-user decodable, provided the stronger users are willing to sacrifice some of their own rate and generate more superfluous codewords.  In other words, we see that users in a set $\Ss$ are further protected from the eavesdropper by the fact that users in set $\Ssc$ are also undecodable, compared to the single-user case.  The TDMA region, on the other hand, does not allow users to help each other this way.  As such, only users whose channel gains allow them to achieve secrecy on their own are allowed to transmit.

For the special degraded case of $h_1 =\dotsc=h_K\triangleq h \le 1$, the perfect secrecy rate region for $\Rsec_k$ becomes the region given by \cite[Theorem 1]{tekin:IT06a} for $\delta=1$.  We also observe that even though there is a limit on the secrecy sum-rate achieved by our scheme, it is possible to send open messages to the intended receiver at rates such that the sum of the secrecy rate and open rate for all users is in the capacity region of the MAC channel to the intended receiver.  Even though we cannot send at capacity with secrecy, the codewords used to confuse the eavesdropper may be used to communicate meaningful information to the intended receiver.

\subsection{The Gaussian Two-Way Wire-Tap Channel}
\label{sec:TW}

In this section, we present an achievable region for the GTW-WT using a superposition coding similar to that used to achieve the region $\GsMACS$ for the GGMAC-WT.  We first define
\begin{definition}[GTW-WT Superposition region, $\GsTW(\Pm)$]
Let $X_k \isnormal{0,P_k}$.  Then, the GTW-WT superposition region, $\GsTW(\Pm)$, is given by
\begin{equation}
\hspace{-\multlinegap}%
\begin{split}
\GsTW(\Pm)= \biggl \{ \Rm \colon \hspace{-.9in} &\\
	&\Rsec_k+\Rpub_k \le I(X_k;Y|X_{k^c}), \quad k=1,2\\
	&\sum_{k \in \Ss} \Rsec_k {\le}
		\biggl[ \sum_{k \in \Ss} I(X_k;Y|X_{k^c}) {-} I(X_\Ks;Z) \biggr]^+ \!\!,
		\, \forall \Ss {\subseteq} \Ks
	\biggr \}
\end{split}
\end{equation}
which can be written as
\begin{equation}
\label{eqn:TWachP}
\hspace{-\multlinegap}%
\begin{split}
\GsTW(\Pm)= \biggl \{ \Rm \colon \hspace{-.9in} &\\
	&\Rsec_k+\Rpub_k \le \onehalf \log \paren{1{+}P_k}, \quad k=1,2\\
	&\sum_{k \in \Ss} \Rsec_k \le \onehalf 
		\biggl[ \sum_{k \in \Ss} \log \paren{1{+}P_k} \\
		&\hspace{.8in} {-}\log \paren{1{+}\frac{\sum_{k \in \Ss}P_k}
			{1{+}\sum_{k \in \Ssc}P_k}} \biggr]^+ \!\!,
		\, \forall \Ss {\subseteq} \Ks
	\biggr \}. \!
\end{split}
\end{equation}
\end{definition}
\begin{remark}
We can also write this region more compactly as the following:
\begin{equation}
\begin{split}
\GsTW(\Pm)= \biggl \{ \Rm \colon \hspace{-.85in} &\\
	&\Rsec_k+\Rpub_k \le \CM_k(\Pm), \quad k=1,2\\
	&\sum_{k \in \Ss} \Rsec_k \le 
		\biggl [\sum_{k \in \Ss} \CM_k(\Pm)-\CWs_\Ss (\Pm) \biggr]^+\!,
		\quad \forall \Ss {\subseteq} \Ks
	\biggr \}. 
\end{split}
\end{equation}
\end{remark}

\begin{theorem}
\label{thm:TWach}
The rate region given below is achievable for the GTW-WT:
\begin{equation}
\text{convex closure of } \bigcup_{\Pm \in \Ps} \GsTW(\Pm).
\label{eqn:TWach}
\end{equation}
\end{theorem}
\begin{figure}[t]
\centering
\includegraphics[width=\figsize,height=.8\figsize,angle=0]{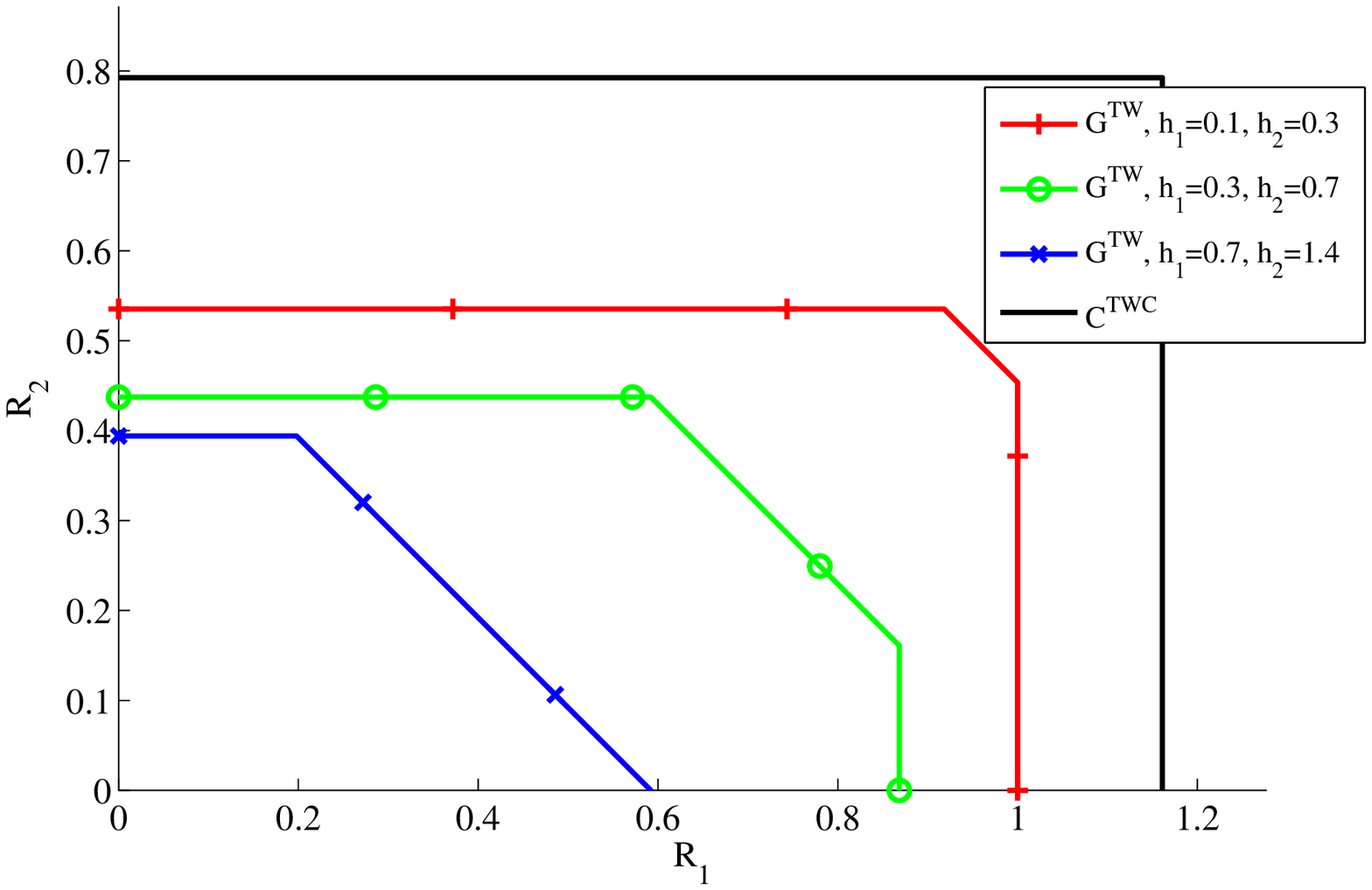}
\caption{\small GTW-WT achievable regions for different channel parameters, $\GsTW(P_1=4, P_2=2)$.}
\label{fig:gtwwtreg}
\vspace{-.2in}
\end{figure}

\begin{IEEEproof}
The proof is very similar to the proof of Theorem \ref{thm:MACach}.  We use the same coding scheme as Theorem \ref{thm:MACach}, the main difference is that we choose the rates to satisfy
\begin{align}
\hspace{-\multlinegap}%
&\Rsec_k+\Rpub_k+\Rx_k \le \onehalf \log \paren{1+P_k}, \quad k=1,2 \\
&\sum_{k\in \Ss} \paren{\Rpub_k+\Rx_k} 
	\le \onehalf \log \paren{1{+}\sum_{k \in \Ss} h_k P_k}, 
	\, \forall \Ss {\subseteq} \Ks \notag \\
	&\hspace{1.8in} \text{with equality if } \Ss=\Ks\\
&\sum_{k \in \Ss} \Rsec_k \le \onehalf
	\biggl[\sum_{k \in \Ss} \log \paren{1+P_k} \notag \\
	&\hspace{.9in} {-}\log \paren{1{+}\frac{\sum_{k \in \Ss}P_k}{1{+}\sum_{k \in \Ssc}P_k}}
	\biggr]^+ \!, \; \forall \Ss {\subseteq} \Ks
\end{align}
or equivalently
\hspace{-\multlinegap}%
\begin{align}
&\label{eqn:TWachcolM} \Rsec_k{+}\Rpub_k{+}\Rx_k \le \CM_k, \quad k=1,2 \\
&\label{eqn:TWRran} \ssum_{k\in \Ss} \paren{\Rpub_k{+}\Rx_k} \le \CW_\Ss, 
	\; \forall \Ss {\subseteq} \Ks,\, \text{with equality if } \Ss{=}\Ks \!\\
&\label{eqn:TWcolsec} \ssum_{k \in \Ss} \Rsec_k 
	\le \bigl[ \ssum_{k \in \Ss} \CM_k -\CW_\Ss \bigr]^+,
	\quad \forall \Ss \subseteq \Ks.
\end{align}
assuming \eqref{eqn:TWcolsec} is positive.  The decodability of $\Wmsec_\Ks$ from $\Ym_1,\Ym_2$ comes from \eqref{eqn:TWachcolM} and the capacity region of the Gaussian Two-Way Channel \cite{han:twoway}. This gives the first set of terms in the achievable region.  The key here is that since each transmitter knows its own codeword, it can \ital{subtract its self-interference} from the received signal and get a clear channel.  Therefore, the Gaussian two-way channel decomposes into two parallel channels.  

The second group of terms in \eqref{eqn:TWachP}, resulting from the secrecy constraint, can be shown the same way as the proof of Theorem \ref{thm:MACach}, since $\Zm$ has the same form for both channels.  In other words, as far as the eavesdropper is concerned, the channel is still a GMAC with $K=2$ users. As such, we need to send $\CW_\Ks$ extra codewords in total, which need to be shared by the two-terminals provided they are not single-user decodable.
\end{IEEEproof}
\begin{figure}[t]
\centering
\includegraphics[width=\figsize,height=.8\figsize,angle=0]{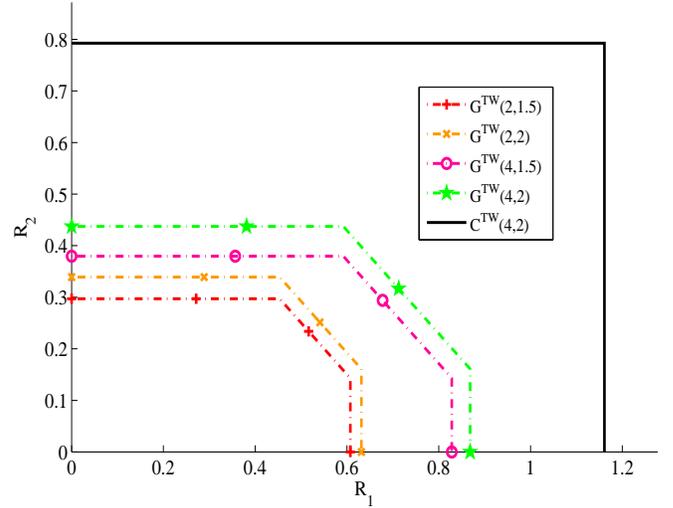}
\caption{\small GTW-WT achievable secrecy region when $\Pmax_1=4, \Pmax_2=2, h_1=.3, h_2=.7$.}
\label{fig:gtwwtUreg}
\vspace{-.2in}
\end{figure}
For different channel gains, the region of all $\Rmsec$ satisfying \eqref{eqn:TWachP} is shown in Figure \ref{fig:gtwwtreg}.  Since we require four dimensions for an accurate depiction of the complete rate region, we only focus on our main interest, i.e., the secrecy rate region.  Figure \ref{fig:gtwwtUreg} shows the achievable secrecy rate region as a function of transmit powers. We note that higher powers always result in a larger region. We indicate the constraint on the overall rates, corresponding to the capacity region of the Gaussian Two-Way Channel, by the dotted line.  Note that the secrecy region has a structure similar to the GGMAC-WT with $K=2$.  As far as the eavesdropper is concerned, there is no difference between the two channels.  However, since the main channel between users decomposes into two parallel channels, higher rates can be achieved between the legitimate terminals (users).  Thus, in effect, each user's transmitted codewords act as a \ital{secret key} for the other user's transmitted codewords, requiring fewer extraneous codewords overall to confuse the eavesdropper, and a larger secrecy region.  We note that a user may either achieve secrecy or not, depending on whether it is single-user decodable or not. As a result, TDMA does not enlarge the region, since each user can at least achieve their single-user secrecy rates.  To see this, note that the constraint on the secrecy sum-rate can be written as:
\begin{align}
\hspace{-\multlinegap}%
\log(1+P_1)+\log(1+P_2)-\log(1+h_1P_1+h_2P_2) \hspace{-2.6in}\notag & \\
	&=\log(1+P_1)-\log(1+h_1P_1) \notag \\
		&\hspace{.5in} +\log(1+P_2) -\log\paren{1+\frac{h_2P_2}{1+h_1P_1}} \\
	&\ge \log(1+P_1)-\log(1+h_1P_1) \notag \\
		&\hspace{.5in} +\log(1+P_2) -\log(1+h_2P_2)
\end{align}
so that transmitting in the two-way channel always provides an advantage over the single-user channels.
\section{Maximization of Sum Rate}
\label{sec:summax}
The achievable regions given in Theorems \ref{thm:MACach} and \ref{thm:TWach} depend on the transmit powers.  We are, thus, naturally interested in the power allocation $\Pmopt$ that would maximize the total secrecy sum-rate.  Recall that the standardized channel gain for user $k$ is $h_k=\frac{\hW_k \nvar_\Mch}{\hM_k \nvar_\Wch}$, and that the higher $h_k$ is, the better the corresponding eavesdropper channel. Without loss of generality, assume that users are ordered in terms of increasing standardized eavesdropper channel gains, i.e., $h_1 \le \dotsc \le h_K$.  Note that, we only need to concern ourselves with the case $h_1 < \dotsc < h_K$, since we can combine users with the same channel gains into one super-user.  We can then split the resulting optimum power allocation for a super-user among the actual constituting users in any way we choose, since they would all result in the same sum-rate.  In addition, from a physical point of view, assuming that the channel parameters are drawn according to a continuous distribution and then fixed, the probability that two users would have the same exact standardized channel gain is zero.

\subsection{GGMAC-WT}

We first examine the superposition region given in \eqref{eqn:MACachP}.  The secrecy sum-rate achievable with superposition coding for the GGMAC-WT was given in Theorem \ref{thm:MACach} as
\begin{equation}
\RsumMACS {=}
\onehalf \mmax{\log \paren{1{+}\sum_{k=1}^K P_k}
	{-} \log \paren{1{+}\sum_{k=1}^K h_k P_k}} \!\!
\end{equation}
and we would like to find the power allocation that maximizes this quantity.  Stated formally, we are interested in the transmit powers that solve the following optimization problem:
\begin{align}
\max_{\Pm \in \Ps} \onehalf \bracket{\log \paren{1+\ssum_{k=1}^K P_k} 
		-  \log \paren{1+\ssum_{k=1}^K h_k P_k}} \notag \hspace{-1.5in} &\\
	&=\min_{\Pm \in \Ps} \; \onehalf \log \hMA_\Ks(\Pm) \\
	&\equiv \min_{\Pm \in \Ps} \; \hMA_\Ks(\Pm) \label{eqn:MACsumprob1}
\end{align}
where 
\begin{equation}
\label{eqn:rhoMACdef}
\hMA_\Ss(\Pm)\triangleq \frac{1+\sum_{k \in \Ss} h_k P_k}{1+\sum_{k \in \Ss} P_k}, \quad 
	 \Ss \subseteq \Ks
\end{equation}
and $\Ss=\Ks$ yields \eqref{eqn:MACsumprob1}.  In obtaining \eqref{eqn:MACsumprob1}, we simply used the monotonicity of the $\log$ function.  The solution to this problem is given below:
\begin{theorem}
\label{thm:sumMAC}
The secrecy sum-rate maximizing power allocation for $\GsMACS$ satisfies $\Popt_k=\Pmax_k$ if $k \le T$ and $\Popt_k=0$ is $k > T$ where $T \in \{0,\dotsc,K\}$ is some limiting user satisfying
\begin{equation}
h_T < \frac{1+\sum_{k=0}^l h_k \Pmax_k}{1+\sum_{k=0}^l \Pmax_k} \le h_{T+1}
\end{equation}
and we define $h_0 \triangleq 0, \, \Pmax_0\triangleq 0$.  Note that this allocation shows that only a \ital{subset of the strong users} must be transmitting.
\end{theorem}
\begin{IEEEproof}
We start with writing the Lagrangian to be minimized:
\begin{equation}
\label{eqn:MACLag}
\Lag(\Pm,\muv) = \hMA_\Ks(\Pm)
	-\sum_{k=1}^K \mu_{1k}P_k + \sum_{k=1}^K \mu_{2k}(P_k-\Pmax_k)
\end{equation}
Equating the derivative of the Lagrangian to zero, we get
\begin{equation}
\label{eqn:MACLagder}
\frac{\del \Lag(\Pmopt,\muv)}{\del \Popt_j} 	= \hdjoptMA -\mu_{1j} + \mu_{2j} = 0
\end{equation}
where we define
\begin{equation}
\label{eqn:rhoMACdotdef}
\hdjMA_\Ss (\Pm) \triangleq \frac{h_j-\hMA_\Ss(\Pm)}{1+\sum_{k\in\Ss} P_k}
\end{equation}
for any set $\Ss \subseteq \Ks$.

It is easy to see that if $h_j > \hoptMA$, then $\mu_{1j}>0$, and we have $\Popt_j=\Pmax_j$.  If $h_j < \hoptMA$, then we similarly find that $\Popt_j=0$.  Finally, if $h_j = \hoptMA$, then we also have 
\begin{equation}
h_j=\frac{1+\sum_{k \in \Ks \setminus j}h_k \Popt_k}{1+\sum_{k \in \Ks \setminus j}h_k \Popt_k}
\end{equation}
and $\hoptMA=\hMA_{\Ks \setminus j}(\Pmopt)$ does not depend on $P_j$, so we can set $\Popt_j=0$ with no effect on the secrecy sum-rate.  Thus, we have $\Popt_j=\Pmax_j$ if $h_j < \hoptMA$, and $\Popt_j=0$ if $h_j \ge \hoptMA$. Then, the optimal set of transmitters is of the form $\Ts=\{1,\dotsc,T\}$ since if a user $T$ is transmitting, all users such that $h_k<h_T$ must also be transmitting. We also note that $\hoptMA=\hMA_\Ts(\Pmmax)$.  Let $T$ be the last user satisfying this property, i.e. $h_T < \hMA_\Ts(\Pmmax)$ and $h_{T+1}\ge\hMA_{\Ts\cup\{T+1\}}(\Pmmax)$.  Note that
\begin{gather}
h_T {<} \frac{1{+}\sum_{k=1}^T h_k \Pmax_k}{1{+}\sum_{k=1}^T \Pmax_k} 
	=\frac{1{+}\sum_{k=1}^{T-1} h_k \Pmax_k {+} h_T \Pmax_T}
		{1{+}\sum_{k=1}^{T-1} \Pmax_k {+} \Pmax_T}\\
h_{T-1} < h_T < \frac{1+\sum_{k=1}^{T-1} h_k \Pmax_k }{1+\sum_{k=1}^{T-1} \Pmax_k}
	=\hMA_{\Ts\setminus\{T\}}(\Pmmax). \label{eqn:hMAT}
\end{gather}

In other words, all sets $\Ss=\{1,\dotsc,S\}$ for $S\le T$ also satisfy this property, and are viable candidates for the optimal set of transmitting users.  Therefore, we can claim that $\Ts$ is the optimum set of transmitting users, since from above we can iteratively see that $\hMA_\Ts(\Pmmax) < \hMA_{\Ss}(\Pmmax)$ for all $S<T$.
\end{IEEEproof}

Note that, for the special case of $K=2$ users, the optimum power allocation is
\begin{equation}
\label{eqn:sumMAC2}
(\Popt_1,\Popt_2) {=} 
	\begin{cases}
	(\Pmax_1,\Pmax_2), & \text{if } h_1 < 1, \, h_2 < \frac{1+h_1 \Pmax_1}{1+\Pmax_1} \\
	(\Pmax_1,0), & \text{if } h_1 < 1, \, h_2 \ge \frac{1+h_1 \Pmax_1}{1+\Pmax_1} \\
	(0,0), & \text{otherwise}
	\end{cases}
\raisetag{2\baselineskip}
\end{equation}

We also need to consider the TDMA region.  In this case, the maximum achievable secrecy sum-rate is:
\begin{equation}
\max_{\substack{0 \le \vg{\alpha} \le 1 \\ \Sigma_{k}\! \alpha_k=1}} \; \sum_{k=1}^K
	\frac{\alpha_k}{2} \bracket{\log \paren{1+\frac{\Pmax_k}{\alpha_k}} 
	-\log\paren{1+\frac{h_k \Pmax_k}{\alpha_k}}}. 
\end{equation}

This is a simple complex optimization problem that can easily be solved numerically.  For the degraded case, we can obtain a closed form solution: $\alpha_k = \frac{\Pmax_k}{\sum_k \Pmax_k}$ as in \cite{tekin:IT06a}.  In general, we cannot obtain such a solution.  However, it is trivial to note that users with $h_k \ge 1$ should not be transmitting in this scheme. The secrecy sum-rate is then the maximum of the solutions given by the superposition and TDMA regions.

\subsection{GTW-WT}
Now, we will examine the power allocation that maximizes the secrecy sum-rate given in Theorem \ref{thm:TWach} as
\begin{equation}
\RsumTW {=} 
\onehalf \bigl[ \log\paren{1{+}P_1}+\log\paren{1{+}P_2} \\
	- \log\paren{1{+}h_1 P_1{+}h_2 P_2} \bigr]^+ \!.
\end{equation}

This problem is formally stated below:
\begin{multline}
\hspace{-\multlinegap}
\max_{\Pm \in \Ps} \! \onehalf \bracket{\log \! \paren{1{+}P_1} 
	{+} \log \! \paren{1{+}P_2} {-} \log \! \paren{1{+}h_1 P_1{+}h_2 P_2}} \\
	\equiv \min_{\Pm \in \Ps} \; \hTW_\Ks(\Pm) 
\label{eqn:TWsumprob1}
\end{multline}
where
\begin{equation}
\label{eqn:hTWdef}
\hTW_\Ss(\Pm) \triangleq \frac{1+\sum_{k \in \Ss} h_k P_k}{\prod_{k \in \Ss}(1+P_k)}
\end{equation}
and $\Ss=\Ks$ yields \eqref{eqn:TWsumprob1}.  The optimum power allocation is stated below:
\begin{theorem}
\label{thm:sumTW}
The secrecy sum-rate maximizing power allocation for the GTW-WT is given by
\begin{equation}
(\Popt_1,\Popt_2) {=} 
	\begin{cases}
	(\Pmax_1,\Pmax_2), &\text{if } h_1 {\le} 1{+}h_2 \Pmax_2,\, h_2 {<} 1{+}h_1 \Pmax_1 \\
	(\Pmax_1,0), &\text{if } h_1 {<} 1,\, h_2 {\ge} 1{+}h_1 \Pmax_1 \\
	(0,0), & \text{otherwise}
	\end{cases}
	\label{eqn:sumTW}
\end{equation}
\end{theorem}
\begin{IEEEproof}
The Lagrangian is,
\begin{equation}
\label{eqn:TWLag}
\Lag(\Pm,\muv) = \hTW_\Ks(\Pm)-\sum_{k=1}^2 \mu_{1k}P_k + \sum_{k=1}^2 \mu_{2k}(P_k-\Pmax_k).
\end{equation}

Equating the derivative of the Lagrangian to zero for user $j$, we get
\begin{equation}
\label{eqn:TWLagder}
\frac{\del \Lag(\Pmopt,\muv)}{\del \Popt_j} 
	= \hdjoptTW -\mu_{1j} + \mu_{2j}  = 0 
\end{equation}
where
\begin{equation}
\label{eqn:rhoTWdotdef}
\hdjTW_\Ks (\Pm) \triangleq \frac{h_j-\frac{1+\sum_{k \in \Ks} h_k P_k}{1+P_j}}{\prod_{k \in \Ks}(1+P_k)}.
\end{equation}

An argument similar to the one for the GGMAC-WT establishes that if $h_j > \paren{1+\sum_{k \in \Ks} h_k P_k} / \paren{1+P_j}$, or equivalently if $h_j > 1+h_\jc \Popt_\jc$, then $\Popt_j = 0$. When equality is satisfied, then $\hdjTW_\Ks(\Pm)=0$ regardless of $P_j$, and as such $\hTW_\Ks(\Pm)$ can be seen to not depend on $P_j$.  To conserve power, we again set $P_j=0$ in this case.  On the other hand, if $h_j < \paren{1+\sum_{k \in \Ks} h_k P_k}/\paren{1+P_j}$, then $\Popt_j=\Pmax_j$.

Consider user 1. If $\Popt_1=0$, and $\Popt_2>0$, this implies that $h_2 < 1$.  Since $h_1 \le h_2 <1$, we cannot have $\Popt_1=0$.  As a consequence of this contradiction, we see that $\Popt_2=0$ whenever $\Popt_1=0$.

Assume $\Popt_1=\Pmax_1$, and consider the two alternatives for $\Popt_2$.  We will have $\Popt_2=0$ if $h_2 \ge 1+h_1 \Pmax_1$; and $\Popt_2=\Pmax_2$ if $h_2<1+h_2 \Pmax_1$.  These cases correspond to $h_1<1$ and $h_1 < 1+h_2 \Pmax_2$, respectively.  Thus, we have \eqref{eqn:sumTW} as the secrecy sum-rate maximizing power allocation.
\end{IEEEproof}

\begin{remark}
Observe that the solution in Theorem \ref{thm:sumTW} has a structure similar to that in Theorem \ref{thm:sumMAC}.  In summary, it is seen that as long as a user is not single-user decodable, it should be transmitting with maximum power.  Hence, when both users can be made to be non-single-user decodable, then the maximum powers will provide the largest secrecy sum-rate.  If this is not the case, then the user who is single-user decodable cannot transmit with non-zero secrecy and will just make the secrecy sum-rate constraint tighter for the remaining user by transmitting open messages.
\end{remark}

Comparing \eqref{eqn:sumTW} to \eqref{eqn:sumMAC2}, we see that the same form of solutions is found, but the range of channel gains where transmission is possible is larger, showing that GTW-WT allows secrecy even when the eavesdropper's channel is not very weak. 

\section{Secrecy Through Cooperative Jamming}
\label{sec:jam}
In the previous section, we found the secrecy sum-rate maximizing power allocations.  For both the GGMAC-WT and GTW-WT, if the eavesdropper is not ``disadvantaged enough" for some users, then these users' transmit powers are set to zero.  We posit that such a user may be able to ``help" a transmitting user, since it can cause more harm to the eavesdropper than to the intended receiver.  We only consider the superposition region, since in the TDMA region a user has a dedicated time-slot, and hence does not affect the others.  We will next show that this type of cooperative behavior is indeed useful, notably exploiting the fact that the established achievable secrecy sum-rate is a difference of the sum-capacity expressions for the intended channel(s) and the eavesdropper's channel.  As a result, reducing the latter more than the former actually results in an \ital{increase} in the achievable secrecy sum-rate.

Formally, the scheme we are considering implies partitioning the set of users, $\Ks$ into a set of transmitting users, $\Ts$ and a set of jamming users $\Tsc = \Ks-\Ts$.  If a user $k$ is jamming, then it transmits $\Xm_k \isnormal{\Popt_k \v{I} ,\v{0}}$ instead of codewords.  In this case, we can show that we can achieve higher secrecy rates when the ``weaker" users are jamming.  We also show that the GTW-WT, has an additional advantage compared to the GGMAC-WT, that is the fact that the receiver already knows the jamming sequence.  As such, this scheme only harms the eavesdropper and not the intended receivers, achieving an even higher secrecy sum-rate.  Once again, without loss of generality, we consider $h_1 < \dotsc < h_K$.  In addition, we will assume that a user can either take the action of transmitting its information or jamming the eavesdropper, but not both.  It is readily shown in Section \ref{sec:jamMAC} below that we do not lose any generality by doing so, and that splitting the power of a user between the two actions is suboptimal from the secrecy sum-rate maximization point of view.

\subsection{GGMAC-WT}
\label{sec:jamMAC}
The problem is formally presented below:
\begin{align}
&\max_{\Ts \subseteq \Ks, \Pm \in \Ps} \;
	\onehalf \biggl [
	\log\paren{1+\frac{\sum_{k \in \Ts} P_k}{1+\sum_{k \in \Tsc} P_k}} \notag \\
		&\hspace{1in} 
		-\log\paren{1+\frac{\sum_{k \in \Ts} h_k P_k}{1+\sum_{k \in \Tsc} h_k P_k}}\biggr]\\
&\hspace{.2in} \equiv \label{eqn:jamMAC}
	\min_{\Ts\subseteq \Ks, \Pm \in \Ps} 
	\frac{\hMA_\Ks(\Pm)}{\hMA_\Tsc (\Pm)}
\end{align}
where we recall that $\hMA_\Ss(\Pm)$ is given  by \eqref{eqn:rhoMACdef}, such that
\vspace{-\abovedisplayshortskip}
\begin{align}
\hMA_\Ks(\Pm) &= \frac{1+\sum_{k \in \Ks} h_k P_k}{1+\sum_{k \in \Ks} P_k} \\
\hMA_\Tsc(\Pm)&= \frac{1+\sum_{k \in \Tsc} h_k P_k}{1+\sum_{k \in \Tsc} P_k}.
\end{align}

To see that a user should not be splitting its power among jamming and transmitting, it is sufficient to note that regardless of how a user splits its power, $\hMA_\Ks(\Pm)$ will be the same, and the user only affects $\hMA_\Tsc (\Pm)$.  Assume the optimum solution is such that user $j$ splits its power, so $j \in \Ts$ and $j \in \Tsc$.  Then, it is easy to see that if $h_j< \hMA_\Tsc (\Pmopt)$, the sum-rate is increased when that user uses its jamming power to transmit, and when $h_j> \hMA_\Tsc(\Pmopt)$, the sum-rate is increased when the user uses its transmit power to jam.  When $h_j = \hMA_\Tsc(\Pmopt)$, then regardless of how its power is split, the sum-rate is the same, and we can assume user $j$ either transmits or jams.

Note that we must have $\hMA_\Ks(\Pm) \le \hMA_\Tsc (\Pm)$ to have a non-zero secrecy sum-rate, and $\hMA_\Tsc(\Pm) >1$ to have an advantage over not jamming.  This scheme can be shown to achieve the following secrecy sum-rate:
\begin{theorem}
\label{thm:jamMAC}
The secrecy sum-rate using cooperative jamming is 
\begin{multline}
\RsumMACJ=
\onehalf \log\paren{1+\frac{\sum_{k \in \Ts} \Popt_k}{1+\sum_{k \in \Tsc} \Popt_k}} \\
	-\onehalf \log\paren{1+\frac{\sum_{k \in \Ts} h_k \Popt_k}
		{1+\sum_{k \in \Tsc} h_k \Popt_k}}
\end{multline}
where $\Ts$ is the set of transmitters and the optimum power allocation is of the form
\begin{equation*}
\{
	\underbrace{\underbrace{1,\dotsc,T}_{\Popt=\Pmax} ,\underbrace{T+1,\dotsc,J-1}_{\Popt=0}}
		_{\text{transmitting, i.e.,}\in \Ts},
	\underbrace{\underbrace{J}_{\Popt_J},
		\underbrace{J+1,\dotsc,K}_{\Popt=\Pmax}}
		_{\text{jamming, i.e.,}\in \Tsc}
\}
\end{equation*}
with
\begin{equation}
\label{eqn:PJ}
\Popt_J = \mmax{\Min{\Pmax_J,\frac{-c_2+\sqrt{c_2^2-4c_1c_3}}{2c_1}}}
\end{equation}
and
\begin{align}
\hspace{-\multlinegap}%
c_1 &{=} h_J \biggl( h_J \sum_{k \in \Ts}\Popt_k	-\sum_{k \in \Ts} h_k \Popt_k \biggr) \\
c_2 &{=} h_J \biggl( 2{+}\!\sum_{k \in \Ks \setminus J} \! h_k \Popt_k
		{+}\!\sum_{k \in \Tsc \setminus J} \!h_k \Popt_k \biggr)
		\sum_{k \in \Ts}\Popt_k \notag\\
	&\hspace{.2in}	-h_J \biggl( 2{+}\sum_{k \in \Ks \setminus J} \! \Popt_k
		{+}\sum_{k \in \Tsc \setminus J} \! \Popt_k \biggr) 
		\sum_{k \in \Ts} h_k \Popt_k \\
c_3 &{=} \biggl( 1{+}\!\sum_{k \in \Ks \setminus J} \! h_k\Popt_k \biggr)
		\biggl(1{+}\!\sum_{k \in \Tsc \setminus J} \! h_k\Popt_k \biggr)
		\sum_{k \in \Ts}\Popt_k \notag \\
	&\hspace{.2in} -h_J \biggl(1{+}\!\sum_{k \in \Ks \setminus J} \! \Popt_k \biggr)
		\biggl(1{+}\!\sum_{k \in \Tsc \setminus J} \! \Popt_k \biggr)
		\sum_{k \in \Ts}h_k \Popt_k \!
\end{align}
whenever the positive real root exists, and $0$ otherwise.
\end{theorem}
\begin{IEEEproof}
We first solve the subproblem of finding the optimal power allocation for a set of given transmitters, $\Ts$.  The solution to this will also give us insight into the structure of the optimal set of transmitters, $\Tsopt$.  We start with writing the Lagrangian:
\begin{equation}
\label{eqn:jamMACLag}
\Lag(\Pm,\muv) = \frac{\hMA_\Ks(\Pm)}{\hMA_\Tsc(\Pm)}
	-\sum_{k=1}^2 \mu_{1k}P_k + \sum_{k=1}^2 \mu_{2k}(P_k-\Pmax_k).
\end{equation}

The derivative of the Lagrangian depends on the user:
\begin{align}
\label{eqn:jamMACLagder}
\hspace{-\multlinegap}%
0 &{=} \frac{\del \Lag (\Pmopt,\muv)}{\del \Popt_j} \notag \\ 
&{=}
\begin{cases}
\frac{\hdjoptMA}{\hMA_\Tsc(\Pmopt)}-\mu_{1j}+\mu_{2j}, &\text{if }j {\in} \Ts\\
\frac{\hdjoptMA \hMA_\Tsc(\Pmopt) -\hoptMA\hdjMA_\Tsc(\Pmopt)} 
	{\hMA^2_\Tsc(\Pmopt)}+\mu_{2j}, &\text{if }j {\in} \Tsc
\end{cases}
\end{align}
since a user $j \in \Tsc$ satisfies $\Popt_j>0$, it must have $\mu_{1j}=0$.
	
Consider a user $j \in \Ts$.  The same argument as in the sum-rate maximization proof leads to $\Popt_j=\Pmax_j$ if $h_j < \hoptMA$ and $\Popt_1=0$ if $h_j \ge \hoptMA$.  Now examine a user $j \in \Tsc$.  We can write \eqref{eqn:jamMACLagder} as
\begin{equation}
\frac{\rho_j(\Pmopt)}{\paren{1+\sum_{k \in \Ks} P_k}^2 \paren{1+\sum_{k \in \Tsc} h_k P_k}^2} +\mu_{2j} =0
\end{equation}
where
\begin{multline}
\label{eqn:rhoj}
\hspace{-\multlinegap}
\rho_j(\Pm) \triangleq -h_j \biggl(1{+}\!\sum_{k \in \Ks} P_k \biggr)
		\biggl( 1{+}\!\! \sum_{k \in \Tsc}\! P_k \biggr)\! \sum_{k \in \Ts} h_k P_k \\
	{+} \biggl(1{+}\!\sum_{k \in \Ks} h_k P_k \biggr)
		\biggl(1{+}\!\! \sum_{k \in \Tsc}\! h_k P_k \biggr)\! \sum_{k \in \Ts} P_k.
		\hspace{-\multlinegap} \;
\end{multline}

Let 
\begin{equation}
\hJMA \triangleq \hMA_\Ks(\Pm) \hMA_\Tsc(\Pm) \frac{\sum_{k \in \Ts} P_k}{\sum_{k \in \Ts} h_k P_k}.
\end{equation}

Then, we have $\rho_j(\Pm) \le 0$ iff $h_j \ge \hJMA$, and $\rho_j(\Pm) \ge 0$ iff $h_j \le \hJMA$.  Thus, we again find that we must have $h_j \ge \hJoptMA$ for all $j \in \Tsc$.  Also, if $h_j>\hJoptMA$, then $\Popt_j=\Pmax_j$.  Only if $h_j=\hJoptMA$, can we have $0<\Popt_j<\Pmax_j$.

Now, since $\hMA_\Tsc(\Pmopt) {\ge} \hoptMA$, we must have $\hMA_\Tsc(\Pmopt) \ge \hoptMA \ge \paren{\sum_{k \in \Ts} h_k \Popt_k}/\paren{\sum_{k \in \Ts} \Popt_k}$.  Thus, we find that $\hJoptMA \ge \hoptMA$.  Then, we know that for a given set of transmitters, $\Ts$, the solution is such that all users $j \in \Ts$ transmit with power $\Pmax_j$ if $h_j \le \hoptMA$.  In the set of jammers $\Tsc$, all users have $h_j \ge \hJoptMA \ge \hMA_\Tsc(\Pmopt)\ge \hoptMA$, and when this inequality is not satisfied with equality, the jammers jam with maximum power.  If the equality is satisfied for some users $j$, their jamming powers can be found from solving $h_j = \hJoptMA$.  By rearranging terms in \eqref{eqn:rhoj}, we note that the optimum power allocation for this user, call it user $J$, is found by solving the quadratic 
\begin{equation}
\label{eqn:PJquad}
\rho_J(\Pmopt)=c_1\Popt_J{}^2+c_2 \Popt_J+c_3=0
\end{equation}
the solution of which is given in \eqref{eqn:PJ}.  

Note that \eqref{eqn:PJquad} defines an (upright) parabola.  If the root given in \eqref{eqn:PJquad} exists and is positive, then $\Popt_J=\Min{\Pmax_j,\Popt_J}$.  This comes from the fact that if $\Popt_J>\Pmax_J$, then $\rho_J(\Pm)<0$ for all $0<P_J<\Pmax_J$, and we must have $\mu_{2J}>0$.  If, on the other hand, \eqref{eqn:PJquad} gives a complex or negative solution, then the parabola does not intersect the $P_J$ axis, and is always positive.  Hence, $h_J<\hJoptoptMA$, and $J$ does not belong to $\Tsc$, i.e. $\Popt_J=0$.

The form of this solution is intuitively pleasing, since it makes more sense for ``weaker" users to jam as they harm the eavesdropper more than they do the intended receiver.  What we see is that all transmitting users $j$, such that $\Popt_j>0$, transmit with maximum power as long as their standardized channel gain $h_j$ is less than some limit $\hoptMA$, and all jamming users must have $h_j > \hoptMA$.  

We claim that all users in $\Tsopt$ must have $h_j<\hJoptMA$ and all users in $\Tsoptc$ have $h_j \ge \hJoptMA$.  To make this argument, we need to show that a $\Ts$ such that there exists some $m \in \Ts$ with $\Popt_m=0$ and $n \in \Tsc$ such that $h_m>h_n$ cannot be the optimum set.  To see this, let $\Pmopt$ be the optimum power allocation for a set $\Ts$. Consider a new power allocation and set such that $\Tsalt=\Ts \setminus \{m\}$, i.e., user $m$ is now jamming, and let $\Palt_k=\Popt_k, \forall k \neq m,n$, $\Palt_m=\pi$ and $\Palt_n=\Popt_n-\pi$, for some small $\pi$.  We then have
\begin{align}
\frac{\hMA_\Ks(\Pmalt)}{\hMA_\Tsaltc(\Pmalt)} 
	&=\efrac{\mnormal{\frac{1+\sum_{k \in \Ks} h_k \Palt_k}{1+\sum_{k \in \Ks} \Palt_k}}}
		{\mnormal{\frac{1+\sum_{k \in \Tsaltc} h_k \Palt_k}{1+\sum_{k \in \Tsaltc} \Palt_k}}} \\
	&=\efrac{\mnormal{\frac{1+\sum_{k \in \Ks} h_k \Popt_k+(h_m-h_n) \pi}
		{1+\sum_{k \in \Ks} \Popt_k}}}
		{\mnormal{\frac{1+\sum_{k \in \Tsc} h_k \Popt_k+(h_m-h_n)\pi}
		{1+\sum_{k \in \Tsc} \Popt_k}}} \\
	&<\efrac{\mnormal{\frac{1+\sum_{k \in \Ks} h_k \Popt_k}
		{1+\sum_{k \in \Ks} \Popt_k}}}
		{\mnormal{\frac{1+\sum_{k \in \Tsc} h_k \Popt_k}
		{1+\sum_{k \in \Tsc} \Popt_k}}} \\
	&=\frac{\hoptMA}{\hMA_\Tsc(\Pmopt)}
\end{align}
which is a lower value for the objective function, proving that $\paren{\Ts,\Pmopt}$ is not optimum.  This shows that all users $j \in \Tsoptc$ must have $h_j > h_k$ for all users $k \in \Tsopt$.  Since the last user in $\Tsoptc$ has $h_j = \hJoptMA$, necessarily $h_j \ge \hJoptMA$ for all $j \in \Tsoptc$, and $h_j<\hJoptMA$ for all $j \in \Tsopt$.

Summarizing, the optimum power allocation is such that there is a set of transmitting users $\{1,\dotsc,T\}$ with $\Popt_k=\Pmax_k$ for $k=1,\dotsc,l$, there is a set of silent users $\{T+1,\dotsc,J-1\}$, and a set of jamming users $\{J,\dotsc,K\}$ with $\Popt_k=\Pmax_k$ for $k=J+1,\dotsc,K$ and $\Popt_J$ is found from $h_J=\hJoptMA$.  This is what is presented in the statement in Theorem \ref{thm:jamMAC}.

Note that to find $T,J$, we can simply do an exhaustive search as we have narrowed the number of possible optimal sets to $K(K-1)$ instead of $2^K-1$ and found the optimal power allocations for each.
\end{IEEEproof}
\bold{Two-user GGMAC-WT:}

For illustration purposes, let us consider the familiar case with $K=2$ transmitters. In this case, we know that either user 2 jams, or no user does.  The solution can be found from comparing the two cases.  If, without jamming, user $2$ can transmit, then it is optimal for it to continue to transmit, and jamming will not improve the sum-rate.
Otherwise, user 2 may be jamming to improve the secrecy rate of user 1.  

The optimum power allocation for user 1 is equivalent to $\Popt_1=\Pmax$ if $h_1 < \hoptMA$ and $\Popt_1=0$ if $h_1 \ge \hoptMA$.  The power for user $2$ is found from \eqref{eqn:PJ}.  For $2$ users, we can simply write \eqref{eqn:PJquad} as
\begin{equation}
\Popt_1 h_2 (h_2-h_1) (\Popt_2-p)(\Popt_2-\bar p) =0
\end{equation}
where
\begin{align}
p&=\frac{-h_2(1-h_1) + \sqrt{D}}{h_2(h_2-h_1)}, \\
\bar p &=\frac{-h_2(1-h_1) - \sqrt{D}}{h_2(h_2-h_1)}\\
D&=h_1h_2(h_2-1)\bracket{(h_2-1)+(h_2-h_1)\Popt_1}.
\end{align}
\begin{figure}[t]
\centering
\includegraphics[width=\figsize,height=.8\figsize,angle=0]{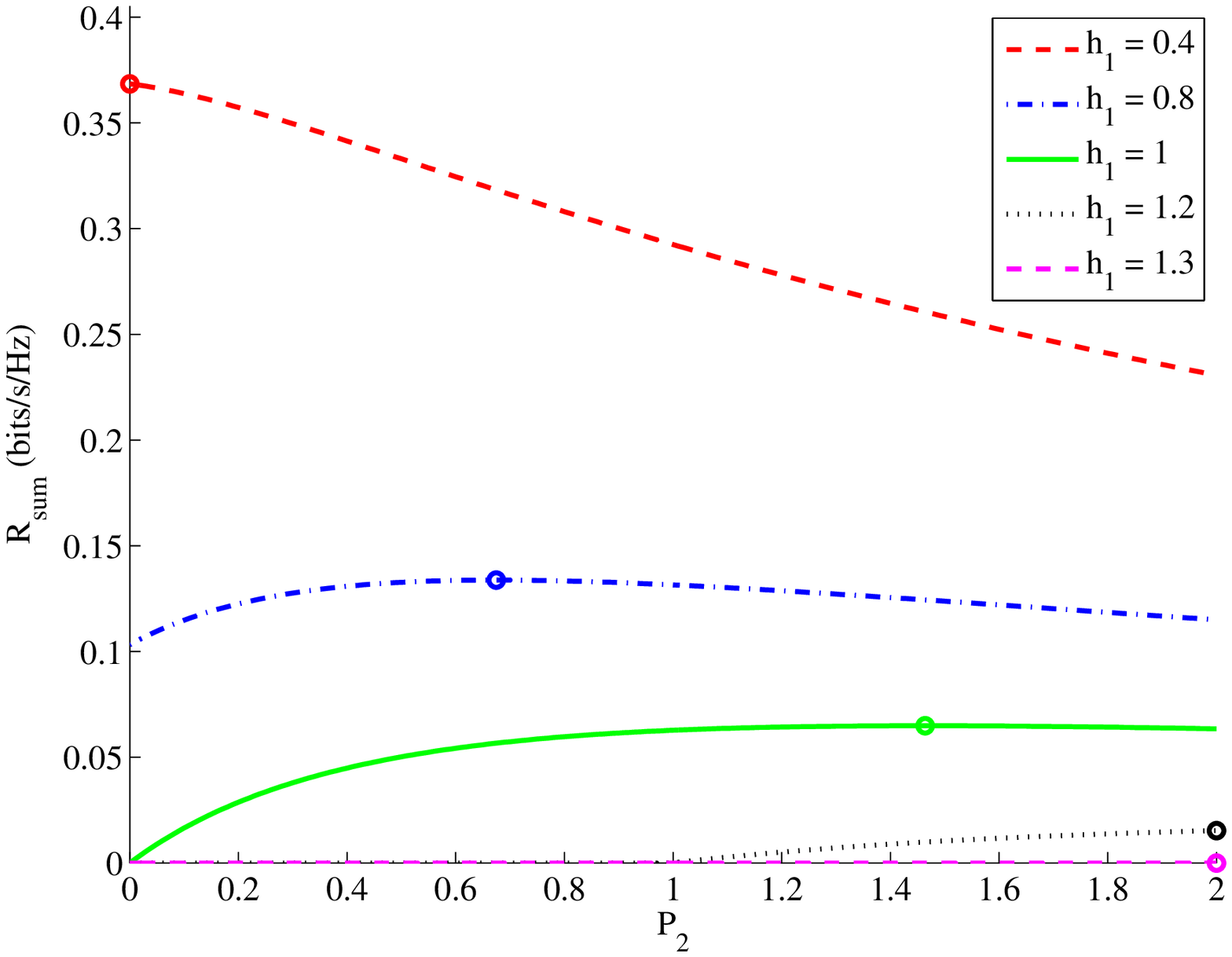}
\caption{\small GGMAC-WT cooperative jamming secrecy sum-rate as a function of $P_2$ with different $h_1$ for $\Pmax_1=\Pmax_2=2, \, h_2=1.4$.  The circles indicate optimum jamming power.}
\label{fig:ggmacwtjamP2}
\vspace{-.2in}
\end{figure}

If $h_1<1$, we automatically have $\Popt_1=\Pmax$.  In addition, we have $\bar p<0$, so we only need to concern ourselves with the possibly positive root, $p$.  We first find when $\Popt_2=0$.  We see that $\rho_2(\Pm) \ge 0$ for all $\Pm \in \Ps$ if $p<0$, equivalent to having two negative roots, or $D<0 \Rightarrow h_2 \le \phi_1(\Pmax)$, equivalent to having no real roots of $\rho_2(\Pm)$.  Now examine when $0<\Popt_2<\Pmax$. This is possible if and only if $\rho_2(\Pmopt)=0$.  Since $\Popt_1>0$, this happens only when $h_1=h_2$ or $\Popt_2=p>0$.  However, if $h_1=h_2$, we are better off transmitting than jamming.  The last case to examine is when $\Popt_2=\Pmax$. This implies that $\rho_2(\Pmmax)<0$, and is satisfied when $p>\Pmax_2$.

Assume $h_2 > h_1 \ge 1$.  In this case, we are guaranteed $p\ge 0$.  If $\Popt_1=0$, then we must have $\Popt_2=0$ since the secrecy rate is 0.  We would like to find when we can have $\Popt_1>0$. Since $h_1 < \phi_2(\Popt_2)$, we must have $\Popt_2 >\frac{h_1-1}{h_2-h_1} \ge 0$, and $\rho_2(\Pmax,\Popt_2) \le 0$.  This implies $\bar p \le \Popt_2 \le p$.  It is easy to see that $\Popt_2=\min \braces{p,\Pmax}$ if $\frac{h_1-1}{h_2-h_1} < \min \braces{p,\Pmax_2}$ and $\Popt_2=0$ otherwise.

Thus, for $K=2$ users, the solution simplifies to:
\begin{multline}
\label{eqn:jamMAC2}
\hspace{-\multlinegap}
(\Popt_1,\Popt_2) =\\ 
\begin{cases}
	(\Pmax_1,0), &\text{if } h_1 {\le} 1, \frac{1+h_1 \Pmax_1}{1+\Pmax_1} {\le} h_2 \le 1\\
	(\Pmax_1,\mmax{\min \braces{p,\Pmax_2}}), &\text{if } h_1 {\le} 1, h_2 {>} 1\\
	(0,0), &\text{if } h_1 {\ge} 1, h_2 {\ge} h_1, \frac{h_1-1}{h_2-h_1}{\ge} \Pmax_1 \\
	(\Pmax_1,\min \braces{p,\Pmax_2}), &\text{if } h_1 {\ge} 1, h_2 {>} h_1, 	
		\frac{h_1-1}{h_2-h_1} {<} \Pmax_1
\end{cases} \hspace{-\multlinegap}
\end{multline}
where 
\begin{equation}
p=\frac{h_1{-}1}{h_2{-}h_1} + \frac{\sqrt{h_1 h_2 (h_2{-}1) \bracket{(h_2{-}1){+}(h_2{-}h_1)\Pmax_1}}}{h_2 (h_2{-}h_1)}.
\end{equation}

This solution can be checked to be in accordance with the sum-rate maximizing power allocation of Theorem \ref{thm:sumMAC}.  We note that in the case unaccounted for in \eqref{eqn:jamMAC2}, i.e., when $h_1 \le 1$ and $h_2 \le \frac{1+h_1\Pmax_1}{1+\Pmax_1}$, both users should be transmitting.  In general, the solution shows that the ``weaker" user should jam if it is not single-user decodable, and if it has enough power to make the other user ``strong" in the new effective channel.
\begin{figure}[t]
\centering
\includegraphics[width=\figsize,height=.8\figsize,angle=0]{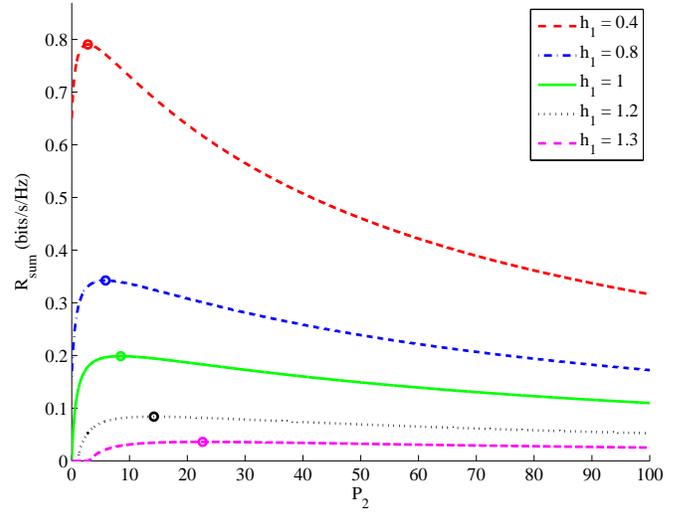}
\caption{\small GGMAC-WT cooperative jamming secrecy sum-rate as a function of $P_2$ with different $h_1$ for $\Pmax_1=\Pmax_2=100, \, h_2=1.4$.  The circles indicate optimum jamming power.}
\label{fig:ggmacwtjamP100}
\vspace{-.2in}
\end{figure}

\subsection{GTW-WT}
\label{sec:jamTW}
Once again, we propose to maximize the secrecy sum-rate using cooperative jamming when useful.  This problem is formally stated as follows:
\begin{multline}
\max_{\substack{\Ts \subseteq \Ks,\\ \Pm \in \Ps}} \onehalf \bracket{
		\sum_{k \in \Ts} \log \paren{P_k}-
		\log \paren{1{+}\frac{\sum_{k \in \Ts} h_k P_k}{1{+}\sum_{k \in \Tsc} h_k P_k}}}\\
\equiv \min_{\Ts \subseteq \Ks} \min_{\Pm \in \Psmax} \frac{\hTW_\Ks(\Pm)}{\hTW_\Tsc(\Pm)}
\label{eqn:jamTW}
\end{multline}
where we recall that $\hTW_\Ss(\Pm)$ is given by \eqref{eqn:hTWdef} and
\begin{align}
\hTW_\Ks(\Pm) &= \frac{1+\sum_{k \in \Ks} h_k P_k}{\prod_{k \in \Ks}(1+P_k)} \\
\hTW_\Tsc(\Pm) &= \frac{1+\sum_{k \in \Tsc} h_k P_k}{\prod_{k \in \Tsc}(1+P_k)}.
\end{align}

Note that $\Ks=\{1,2\}$ since there are only two terminals. A similar argument to the GGMAC-WT case can easily be used to establish that we can assume a user to be either transmitting or jamming, but not both.  Since the jamming user is also the receiver that the other user is communicating with and knows the transmitted signal, this scheme entails no loss of capacity as far as the transmitting user is concerned.  The optimum power allocations are given as follows.
\begin{theorem}
\label{thm:jamTW}
The achievable secrecy sum-rate for the the collaborative scheme described is
\begin{equation}
\RsumTWJ{=}
\sum_{k \in \Ts} \! \onehalf \log \paren{\Popt_k}{-}\onehalf \log \!
	\paren{1{+}\frac{\sum_{k \in \Ts} h_k \Popt_k}{1{+}\sum_{k \in \Tsc} h_k \Popt_k}} 
	\hspace{-.06in}
\end{equation}
where $\Ts$ is the set of transmitting users and the optimum power allocations are given by
\begin{multline}
\hspace{-\multlinegap}
(\Popt_1,\Popt_2) = \\ 
\begin{cases} \!
(\Pmax_1,\Pmax_2), \, \text{both transmit,} &\!\! \text{if } h_1 {<} h_2 {\le} 1\\ \!
(\Pmax_1,\Pmax_2), \, \text{1 transm., 2 jams,} &\!\! \text{if } h_1 {\le} 1 {<} h_2\\ \!
(\Pmax_1,\Pmax_2), \, \text{1 transm., 2 jams,} 
	& \!\! \text{if } 
		\begin{aligned}[t]
		&1{<}h_1{<}1{+}h_2\Pmax_2, \\
	 	&\hTW_2(\Pmmax){>}\hTW_1(\Pmmax)
	 	\end{aligned} \\ \!
(\Pmax_1,\Pmax_2), \, \text{2 transm., 1 jams,} 
	& \!\!\text{if } 
		\begin{aligned}[t]
		&1{<}h_1{<}h_2{<}1{+}h_1\Pmax_1, \\
		&\hTW_1(\Pmmax){>}\hTW_2(\Pmmax) 
		\end{aligned}\\ \!
(0,0), &\!\! \text{otherwise}
\end{cases} \hspace{-\multlinegap}
\end{multline}
\end{theorem}
\begin{IEEEproof}
Similar to the GGMAC-WT, we start with the sub-problem of finding the optimal power allocation given a jamming set.  The Lagrangian is given by
\begin{equation}
\label{eqn:jamTWLag}
\Lag(\Pm,\muv) = \frac{\hTW_\Ks(\Pm)}{\hTW_\Tsc(\Pm)}
	-\sum_{k=1}^2 \mu_{1k}P_k + \sum_{k=1}^2 \mu_{2k}(P_k-\Pmax_k).
\end{equation}

Taking the derivative we have
\begin{align}
\hspace{-\multlinegap}
0 &{=} \frac{\del \Lag (\Pmopt,\muv)}{\del \Popt_j} \notag \\
&{=}\label{eqn:jamTWLagder}
\begin{cases}
\frac{\hdjoptTW}{\hTW_\Tsc(\Pmopt)}-\mu_{1j}+\mu_{2j},  
	&\text{if } j {\in} \Ts\\
\frac{\hdjoptTW \hTW_\Tsc(\Pmopt) -\hoptTW \hdjTW_\Tsc(\Pmopt)}{\hTW^2_\Tsc(\Pmopt)}+\mu_{2j},
	&\text{if } j {\in} \Tsc
\end{cases} \hspace{-\multlinegap}
\end{align}
since a user $j \in \Tsc$ satisfies $\Popt_j>0$, it must have $\mu_{1j}=0$.  

Consider user $j \in \Ts$.  We again argue that if $h_j > \frac{1+\sum_{k \in \Ks} h_k P_k}{1+P_j}$, then $\Popt_j=0$ and if $h_j < \frac{1+\sum_{k \in \Ks} h_k P_k}{1+P_j}$, then $\Popt_j=\Pmax_j$.

Now examine a user $j \in \Tsc$.  It is easy to see that since such a user only harms the jammer, the optimal jamming strategy should have $\Popt_j=\Pmax_j$, i.e., the maximum power.  This can also be seen by noting that \eqref{eqn:jamTWLagder} for this case simplifies to
\begin{equation}
	\frac{-h_j \sum_{k \in \Ts} h_k \Popt_k}
		{\paren{1{+}\sum_{k \in \Tsc} h_k \Popt_k}^2 \paren{\prod_{k \in\Ks}(1{+}\Popt_k)}^2}
	+\mu_{2j}=0
\end{equation}
and hence we must have $\mu_{2j}>0 \Rightarrow \Popt_j=\Pmax_j$ for all $j \in \Tsc$.

The jamming set will be one of $\emptyset,\{1\},\{2\}$, since there is no point in jamming when there is no transmission.  Also, if any of the two users is jamming, by the argument above, $\Popt_j=\Pmax_j$, $j=1,2$.  

We can easily see that jamming by a user $j$ only offers an advantage if $h_j>1$, i.e., $\hTW_\Tsc(\Pm)>1$ iff $h_j>1$ for $j \in \Tsc$.  Thus, when $h_1 < h_2 \le 1$, both users should be transmitting instead of jamming.  However, when any user has $h_j>1$, jamming always does better than the case when both users are transmitting.  In this case, $\hTW_j(\Pmmax) \ge \hTW_\jc(\Pmmax)$ for some user $j$, and the objective function in \eqref{eqn:jamTW} is minimized when this user is jamming, and the other one is transmitting.  If, however, $h_\jc>1+h_j \Pmax_j$, then it will not transmit, and we should not be jamming.  Consolidating all of these results, we come up with the power allocation in in Theorem \ref{thm:jamTW}.
\end{IEEEproof}
\begin{figure}[t]
\centering
\includegraphics[width=\figsize,height=.8\figsize,angle=0]{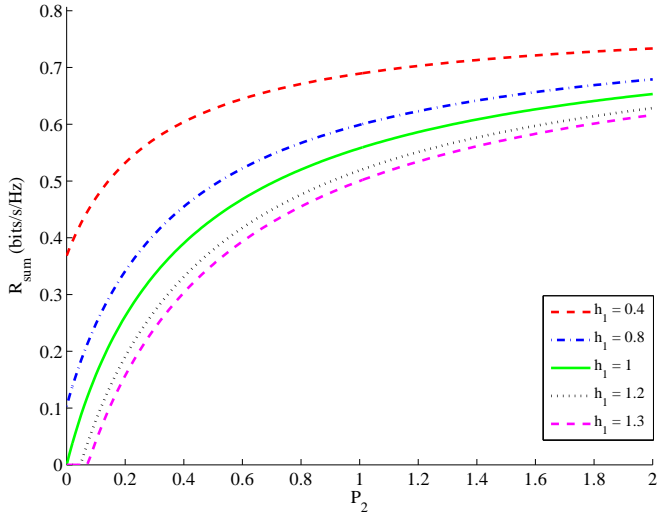}
\caption{\small GTW-WT cooperative jamming secrecy sum-rate as a function of $P_2$ with different $h_1$ for $\Pmax_1=\Pmax_2=2, \, h_2=4.2$.}
\label{fig:twwtjamP2}
\vspace{-.2in}
\end{figure}

\begin{remark}
A sufficient, but not necessary condition for the weaker user to be the jamming user is if $h_2 \Pmax_2 > h_1 \Pmax_1$; this case corresponds to having higher SNR at the eavesdropper for the original, non-standardized model.  This can be interpreted as ``jam with maximum power if it is possible to change user 1's effective channel gain such that it is no longer single-user decodable".  For the simple case of equal power constraints, $\Pmax_1=\Pmax_2=\Pmax$, it is easily seen that user 1 should never be jamming.  The optimal power allocation in that case reduces to
\begin{multline}
(\Popt_1,\Popt_2) = \\
\begin{cases}
(\Pmax,\Pmax), \, \text{both transmit}, & \text{if } h_1 < h_2 \le 1\\
(\Pmax,\Pmax), \, \text{1 transmits, 2 jams}, & \text{if } h_1 < 1+h_2 \Pmax\\
(0,0), & \text{otherwise}
\end{cases}
\end{multline}
\end{remark}

\section{Numerical Results}
\label{sec:results}
In this section, we present numerical results to illustrate the achievable rates obtained, as well as the cooperative jamming scheme and its effect on achievable secrecy sum-rates.

As mentioned earlier in the paper, examples of achievable secrecy rate regions are given in Figures \ref{fig:ggmacwtUreg} and \ref{fig:gtwwtUreg} for the GGMAC-WT with $K=2$ and GTW-WT respectively.  Comparing Figures \ref{fig:ggmacwtUreg} and \ref{fig:gtwwtUreg}, we see that the GTW-WT achieves a larger secrecy rate region then the GGMAC-WT, and offers more protection to ``weak" users.  In addition, TDMA does not enlarge the achievable region for GTW-WT since superposition coding always allows users to achieve their single-user secrecy rates for any transmit power.

Let us have a closer look at the secrecy advantage of the two-way channel over the MAC with two users.  For the GGMAC-WT with $K=2$, the achievable maximum secrecy sum-rate, $\Rsec_1+\Rsec_2$ is limited by the channel parameters.  It was shown in \cite{tekin:ISIT06} that for the degraded case, $h \le 1$, the secrecy sum-capacity, $\CM_\Ks(\Pm)-\CW_\Ks(\Pm)$, is an increasing function of the total sum power, $\Pmax_\Sigma=\Pmax_1+\Pmax_2$.  However, it is limited since $\CM_\Ks(\Pm)-\CW_\Ks(\Pm) \rightarrow -\onehalf \log h$ as $\Pmax_\Sigma \toinf$.  For the general case, where $\Pmax_1,\Pmax_2 \toinf$, Theorem \ref{thm:sumMAC} implies that the sum-rate is maximized when only user 1 transmits (assuming $h_2 > h_1$), and is bounded similarly by $-\onehalf \log h_1$.  On the other hand, For the GTW-WT, unlike the GGMAC-WT, it is possible to increase the secrecy capacity by increasing the transmit powers. This mainly stems from the fact that the users now have the extra advantage over the eavesdropper that they know their own transmitted codewords.  In effect, \ital{each user helps encrypt the other user's transmission}.  To see this more clearly, consider the symmetric case where $\alpha_1=\alpha_2=h_1=h_2=1$ and $\Pmax_1 = \Pmax_2 = \Pmax$, which makes all users receive a similarly noisy version of the same sum-message.  The only disadvantage the eavesdropper has, is that he does not know any of the codewords whereas user $k$ knows $\Xm_k$.  In this case, $\Rsec_1+\Rsec_2 \le \onehalf \log \paren{1+\Pmax^2/(1+2\Pmax)}$ is achievable, and this rate approaches $\onehalf \log (\onehalf \Pmax)$ as $\Pmax \gg 1$.  Thus, it is possible to achieve a secrecy-rate increase at the same rate as the increase in channel capacity.

Next, we examine the secrecy sum-rate maximizing power allocations and optimum powers for the cooperative jamming scheme.  Figures \ref{fig:ggmacwtjamP2} and \ref{fig:ggmacwtjamP100} show the achievable secrecy rate improvement for the cooperative jamming scheme for various channel parameters for the GGMAC-WT with $K=2$.  The plots are the secrecy rates for user 1 when user 2 is jamming with a given power, which correspond to user 1's single-user secrecy \ital{capacity}, \cite{leung-hellman:gaussianwiretap}, since only one user is transmitting.  When $h_1 \ge 1$, the secrecy capacity is seen to be zero, unless user $2$ has enough power to \ital{convert} user 1's re-standardized channel gain to less than 1.  For the GTW-WT, it is always optimal for user $2$ to jam as long as it enables user $1$ to transmit, as seen in Figure \ref{fig:twwtjamP2}.  The results show, as expected, that secrecy is achievable for both users so long as we can keep the eavesdropper from single-user decoding the transmitted codewords by treating the remaining user as noise.

\begin{figure}[t]
\centering
{\includegraphics[width=\figsize,angle=0]{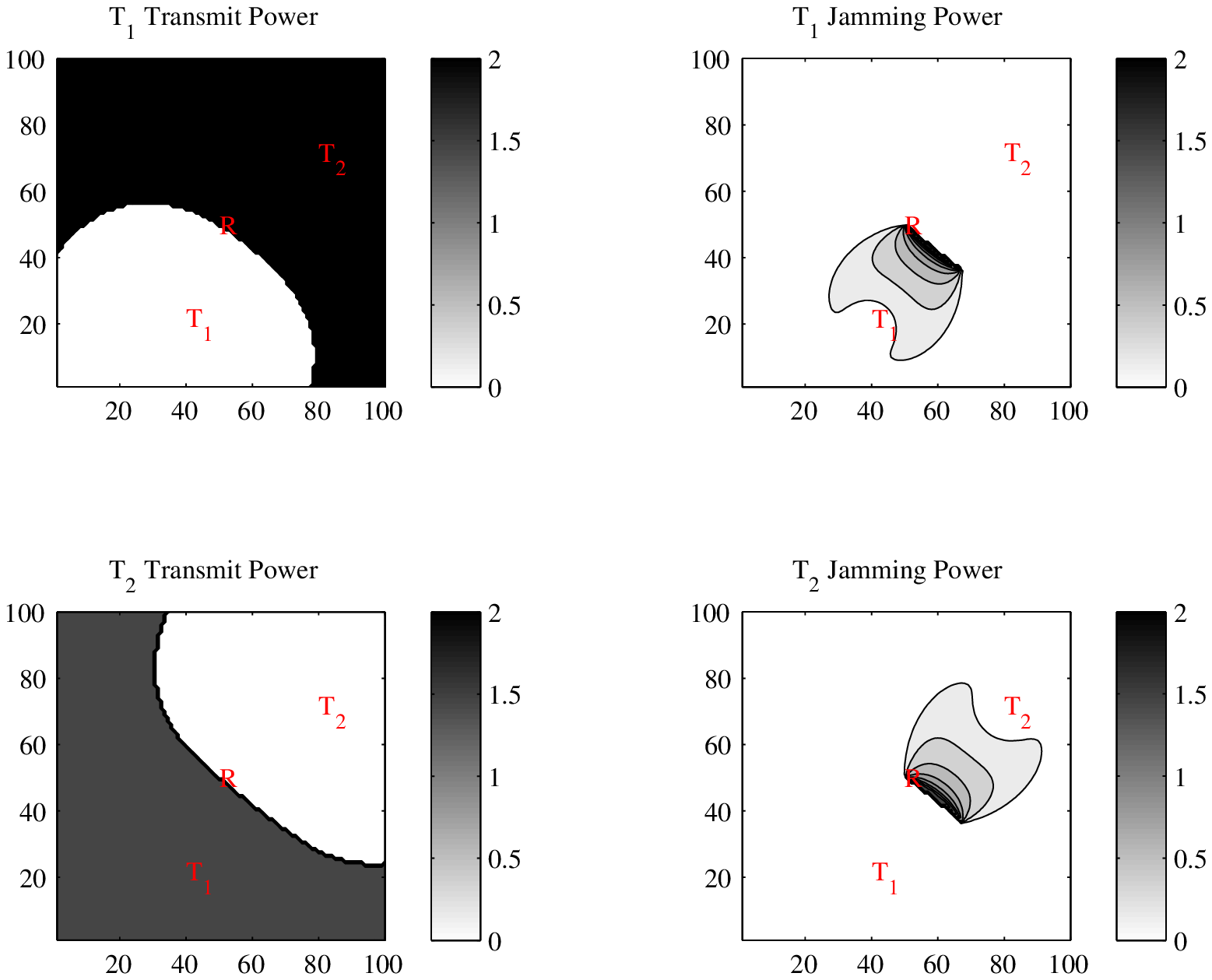}
\vspace{.2in}}
\includegraphics[width=\figsize,angle=0]{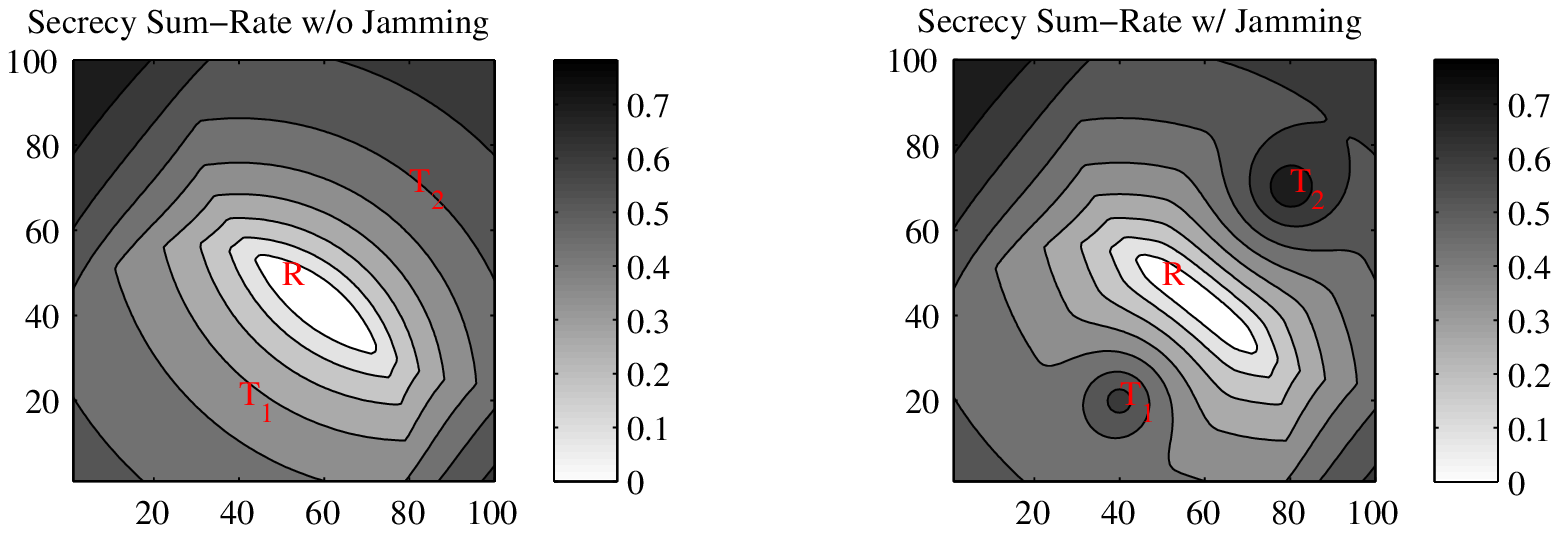}
\caption{\small GGMAC-WT cooperative jamming example - darker shades correspond to higher values.}
\label{fig:ggmacwtcoljam}
\end{figure}

\begin{figure}[t]
\centering
{\includegraphics[width=\figsize,angle=0]{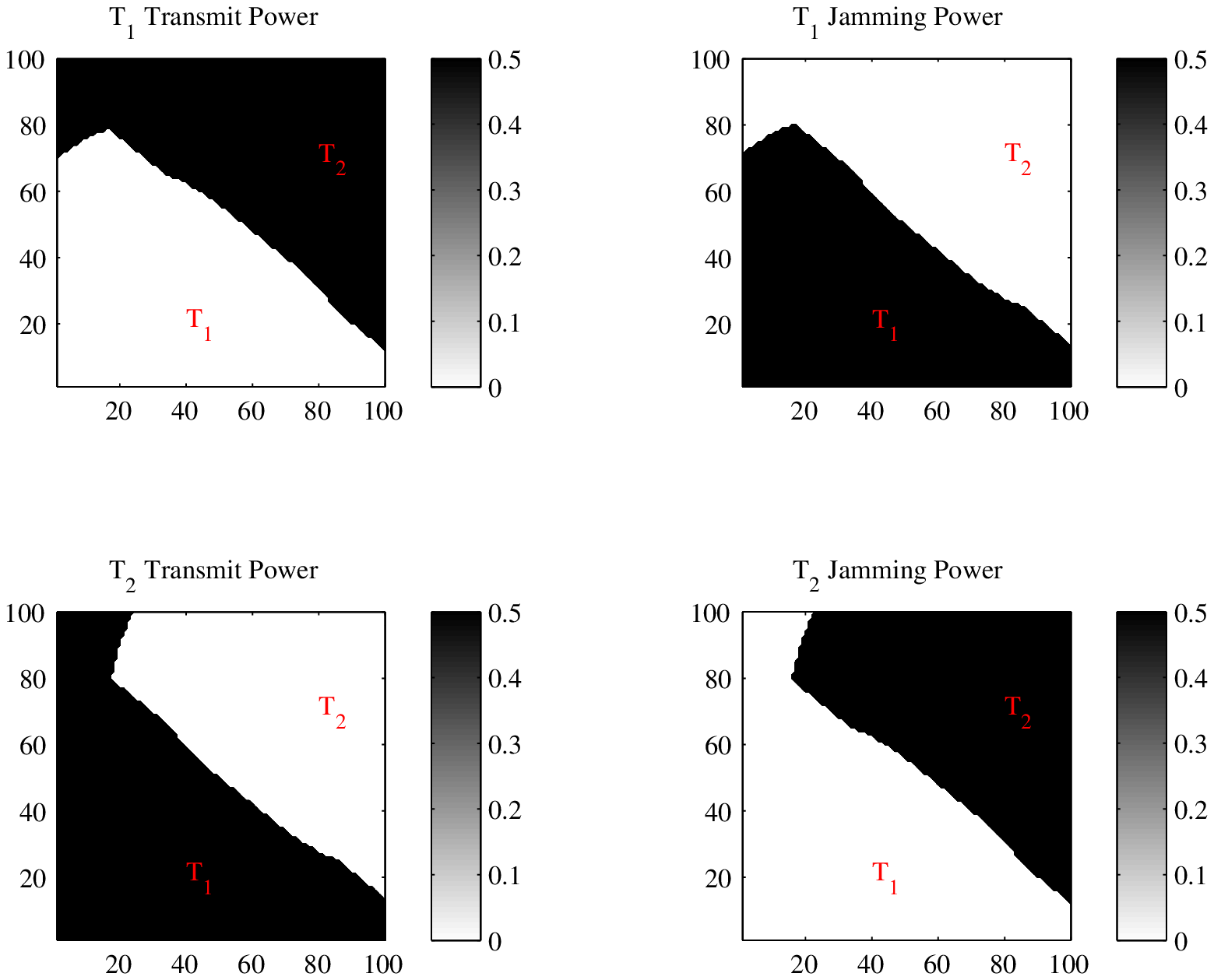}
\vspace{.2in}}
\includegraphics[width=\figsize,angle=0]{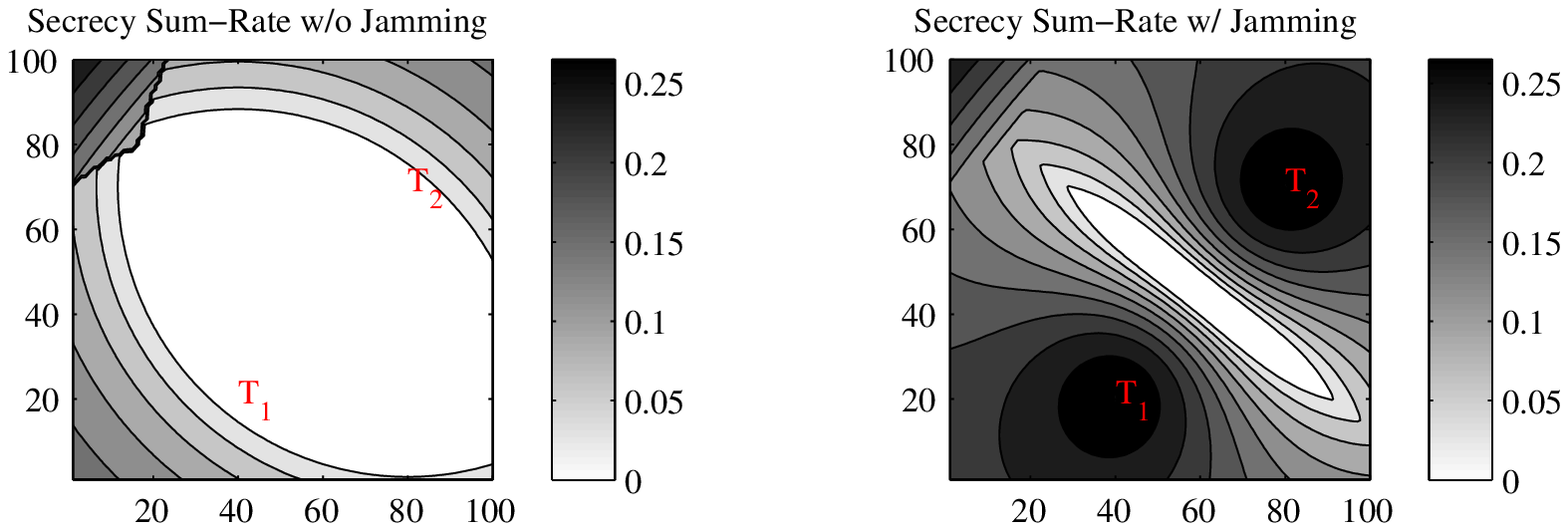}
\caption{\small GTW-WT cooperative jamming example - darker shades correspond to higher values.}
\label{fig:gtwwtcoljam}
\end{figure}

Since the coding schemes considered here assume knowledge of eavesdropper's channel gains, applications are limited.  One practical application could be securing of a physically protected area such as inside a building, when the eavesdropper is known to be outside.  In such a case we can design for the worst case scenario.  An example is given in Figure \ref{fig:ggmacwtcoljam} for the GGMAC-WT, where we assume a simple path-loss model and fixed locations for two transmitters (T) and one receiver (R) at the center. We examine the transmit/jam powers for this area when the eavesdropper is known to be at $(x,y)$ using a fixed path-loss model for the channel gains, and plot the transmit/jam powers and the achieved secrecy sum-rates as a function of the eavesdropper location.  It is readily seen that when the eavesdropper is close to the BS, the secrecy sum-rate falls to zero.  Also, when the eavesdropper is in the vicinity of a transmitter, that transmitter cannot transmit in secrecy.  However, in this case, the transmitter can jam the eavesdropper effectively, and allow the other transmitter to transmit and/or increase its secrecy rate with little jamming power.  The situation for the GTW-WT is similar, and is shown in Figure \ref{fig:gtwwtcoljam}.  In this case, jamming is more useful as compared to the GGMAC-WT, and we see that it is possible to provide secrecy for a much larger area where the eavesdropper is located, as the jamming signal does not hurt the intended receiver.

\section{Conclusions and Future Work}
\label{sec:conclusion}
In this paper, we have considered the Gaussian multiple access and two-way channels in the presence of an external eavesdropper who receives the transmitted signals through a multiple-access channel, and provided achievable secrecy rates.  We have shown that the multiple-access nature of the channels considered can be utilized to improve the secrecy of the system.  In particular, we have shown that the total extra randomness is what matters mainly concerning the eavesdropper, rather than the individual randomness in the codes.  As such, it may possible for users whose single-user wire-tap capacity are zero, to communicate with non-zero secrecy rate, \ital{as long as it is possible to put the eavesdropper at an overall disadvantage}.  This is even clearer for two-way channels, where even though the eavesdropper's channel gain may be better than a terminal's, the extra knowledge of its own codeword by that terminal enables communication in perfect secrecy as long as the eavesdropper's received signal is not strong enough to allow single-user decoding.

We found achievable secrecy rate regions for the General Gaussian Multiple-Access Wire-Tap Channel (GGMAC-WT) and the Gaussian Two-Way Wire-Tap Channel (GTW-WT).  We also showed that for the GGMAC-WT the secrecy sum-rate is maximized when only users with `strong" channels to the intended receiver as opposed to the eavesdropper transmit, and they do so using all their available power.  For the GTW-WT, the sum-rate is maximized when both terminals transmit with maximum power as long as the eavesdropper's channel is not good enough to decode them using single-user decoding.

Finally, we proposed a scheme termed \ital{cooperative jamming}, where a disadvantaged user may help improve the secrecy rate by jamming the eavesdropper.  We found the optimum power allocations for the transmitting and jamming users, and showed that significant rate gains may be achieved, especially when the eavesdropper has much higher SNR than the receivers and normal secret communications is not possible.  The gains can be significant for both the GGMAC-WT and GTW-WT.  This cooperative behavior is useful when the maximum secrecy sum-rate is of interest.  We have also contrasted the secrecy rates of the two channels we considered, noting the benefit of the two-way channels where the fact that each receiver has perfect knowledge of its transmitted signal brings an advantage with each user effectively encrypting the communications of the other user.

In this paper, we only presented achievable secrecy rates for the GGMAC-WT and GTW-WT.  The secrecy capacity region for these channels are still open problems. In \cite{tekin:Allerton07}, we also found an upper bound for the secrecy sum-rate of the GGMAC-WT and noted that the achievable secrecy sum-rate and the upper bound we found only coincide for the degraded case, so that we have the secrecy sum-capacity for the degraded GMAC-WT. Even though there is a gap between the achievable secrecy sum-rates and upper bounds, cooperative jamming was shown in \cite{tekin:Allerton07} to give a secrecy sum-rate that is close to the upper bound in general.

Finally, we note that the results provided are of mainly theoretical interest, since as of yet there are no currently known practical codes for multi-access wire-tap channels unlike the single-user case where in some cases practical codes have been shown to be useful for the wire-tap channel, \cite{thangarajetal:ldpcwiretap,blochetal:wirelesssec2}. Furthermore, accurate estimates of the eavesdropper channel parameters are required for code design for wire-tap channels where the channel model is quasi-static, as in our models considered in this paper.
\IEEEtriggeratref{41}


\end{document}